\documentclass[fleqn,usenatbib]{mnras}

\usepackage[T1]{fontenc}
\usepackage{ae,aecompl}
\usepackage{ulem}

\usepackage{graphicx}	
\usepackage{amsmath}	

\usepackage{amssymb}	
\usepackage{caption}
\usepackage{siunitx}
\usepackage{hyperref}
\usepackage{upgreek}
\usepackage{subfig, float}
\usepackage{newtxtext,newtxmath}

\newcommand{\minute}{$^\prime$}
\DeclareSIUnit\arcsec{''}

\title[Rotation periods of ultra-cool dwarfs]{The photometric periods of rapidly rotating field ultra-cool dwarfs}

\author[P. A. Miles-P\'aez et al.]{
Paulo A. Miles-P\'aez,$^{1,2}$\thanks{pamiles@cab.inta-csic.es}
Stanimir A. Metchev$^{3,4}$ and 
Benjamin George$^{3}$
\\
$^{1}$European Southern Observatory, Karl-Schwarzschild-Stra{\ss}e 2, 85748 Garching, Germany\\
$^{2}$Centro de Astrobiolog\'ia (CSIC-INTA), Carretera de Ajalvir km 4, 28850 Torrej\'on de Ardoz, Madrid, Spain\\
$^{3}$Department of Physics \& Astronomy and Institute for Earth and Space Exploration, Western University, London, ON N6A 3K7, Canada\\
$^{4}$Department of Astrophysics, American Museum of Natural History, Central Park West at 79th Street, New York, NY 10024-5192, USA\\
}

\date{Accepted XXX. Received YYY; in original form ZZZ}

\begin{document}
\label{firstpage}
\pagerange{\pageref{firstpage}--\pageref{lastpage}}
\maketitle

\begin{abstract}

We use 1-m class telescopes and the Transiting Exoplanet Survey Satellite (TESS) to explore the photometric variability of all known rapidly rotating ($v\sin{i}\gtrsim30$ km\,s$^{-1}$) ultra-cool ($\geq$M7) dwarfs brighter than $I\approx17.5$ mag. For a sample of 13 M7--L1.5 dwarfs without prior photometric periods, we obtained $I$-band light curves with the SMARTS 1.3m and WIYN 0.9m telescopes and detected rotation-modulated photometric variability in three of them. Seven of our targets were also observed by TESS and six of them show significant periodicities compatible with the estimated rotation periods of the targets. We investigate the potential of TESS to search for rotation-modulated photometric variability in ultra-cool dwarfs and find that its long stare enables $<$80~h periodic variations to be retrieved with $\leq$1\% amplitudes for ultra-cool dwarfs up to a TESS magnitude of 16.5. We combine these results with the periods of all other known photometrically-periodic ultra-cool dwarfs from the literature, and find that the periods of ultra-cool dwarfs range between 1 and 24 h, although the upper limit is likely an observational bias. We also observe that the minimum rotation periods follow a lower envelope that runs from $\approx$2 h at spectral type $\approx$M8 to $\approx$1 h at spectral type T.\\

\end{abstract}




\section{Introduction}
Stellar spectro-photometric variability provides valuable information on the physical processes that take place at different atmospheric heights of a star. In the case of very low-mass stars and brown dwarfs (usually referred to as ultra-cool dwarfs; spectral types later than M7), spectro-photometric variability can have multiple origins, such as magnetic processes \citep[e.g., flares or Sun-like spots; ][]{1990ApJS...74..225H}, condensates particles that form atmospheric cloud-like structures \citep{1996A&A...308L..29T}, or even brightness changes due to an atmosphere out of chemical equilibrium \citep{2015ApJ...804L..17T}. 

Independent of the variability nature, the monitoring of these brightness changes (usually at red-optical and infrared wavelengths) allows us to derive the rotation periods of ultra-cool dwarfs. Numerous searches for photometric variability have been carried out from both the ground \citep[e.g.,][]{1999MNRAS.304..119T,2005MNRAS.360.1132K,2013ApJ...779..101H,2014ApJ...793...75R,2017MNRAS.472.2297M} and space \citep[e.g.,][]{2014ApJ...782...77B,2015ApJ...799..154M,2016ApJ...823..152C,2019ApJ...883..181M,2021AJ....161..224T,2021A&A...651L...7M,2022ApJ...924...68V}, covering from late-M to Y dwarfs of different ages. Understanding the distribution of rotation periods in ultra-cool dwarfs is crucial for investigating their angular momentum evolution, which seems different from the well-known spin-down for higher mass stars caused by angular momentum loss via magnetized stellar winds \citep{1972ApJ...171..565S}. For example, \citet{2021AJ....161..224T} reported the discovery of three L and T brown dwarfs with rotation periods close to 1 hour, which rather than a spin-down have likely experienced an increase in spin rate: the likely result of evolutionary contraction without disk- or wind-induced angular momentum loss. When comparing to all 78 other L, T, and Y dwarfs with a measured rotation periods, \citet{2021AJ....161..224T} concluded that the 1 hour period marks an empirical lower limit to the spin period of brown dwarfs. This is a factor of two to three slower than the rotational stability limit dictated by angular momentum conservation at substellar masses. The reason for the discrepancy likely lies in stability considerations related to the high oblateness of such rapid rotators \citep{1964ApJ...140..552J}. \citet{2021AJ....161..224T} further invoke the unknown role of the magnetic dynamo of the metallic hydrogen interior, which may be an important contributor to the angular momentum budget at high spin rates. The situation for low-mass stars just above the hydrogen burning limit may be different still, as thermonuclear reactions are an important contributor to the energy balance.

To augment our understanding of rapidly rotating ultra-cool dwarfs, we seek to expand their sample with objects that are already known to have high projected rotational velocities, but lack photometric period measurements. Specifically, we focus on field M/L transition dwarfs with $v \sin i \gtrsim 30$~km s$^{-1}$ that are bright enough to be observed with 1~m-class telescopes at red optical wavelengths. We conducted a ground-based photometric monitoring campaign of 13 such targets with 1~m-class telescopes in 2018. The sample selection, ground-based observing campaign, and data reduction are presented in Sections \ref{sec:sample} and \ref{sec:obs}. With the release of data from the Transit Exoplanet Survey Satellite \citep[TESS;][]{2014SPIE.9143E..20R}, we complemented these with TESS light curves where available. The analysis of the ground-based and TESS observations is explained in Sections \ref{sec:anl} and \ref{sec:spa}, respectively. The TESS data  refined periods for three of our ground-based targets, and revealed periods for three more. We further combine the results from our variability survey with published photometric periods of 15 additional ultra-cool dwarfs, which we also refine with data from TESS, and discuss the ensemble findings in Section \ref{disc}. Overall, the TESS data yield significantly more precise periods, and hence the Sections of greatest interest are \ref{sec:spa} and \ref{disc}. Nevertheless, the ground-based observations do allow us to lift the ambiguity on the component periods of one triple system that is unresolved in TESS.

\begin{table}
\caption{Target sample characteristics.}
\hspace*{-0.28cm}\begin{tabular}{lcccr}
\hline
Target Name & SpT     &TESS &  $v\sin{i}$  & Ref.  \\
 &  & (mag) & (km\,s$^{-1}$) & \\
\hline
2MASS J00242463-0158201	 & M9.5 	 &  15.053 & 33$\pm$3 	&1, 2  \\
LP213-67 &   M7 &  13.549         & 35$\pm$5        &2               \\
LP213-68AB & M8+L0     & 15.069      & 35$\pm$5         &2             \\
2MASS J11593850+0057268 &  L0       & 17.491    & 74.5$^{+9}_{-5.4}$  & 3            \\
2MASS J13365044+4751321 & M7      & 14.987   & 30$\pm$5    &2  \\
2MASS J14112131-2119503 & M7.5    & 14.905     & 44.0$\pm$4.0         &1                   \\
2MASS J14310126-1953489 &  M9     & 18.024     & 47.4$^{+5.2}_{-3.3}$    & 4          \\
LP859-1     &  M7    & 14.454      & 30$\pm$5  &2   \\
2MASS J16192988-2440469 &  M8     & 17.044      & 47$\pm$2    &5             \\
2MASS J17054834-0516462 &  L1     & 16.688   & 26$\pm$2.6   &6          \\
2MASS J20575409-0252302 &  L1.5    & 16.606       & 62$\pm$6.2       & 3                 \\
2MASS J23515044-2537367 & M9      & 15.608      & 36$\pm$4        &1, 3 \\
\hline
\multicolumn{5}{p{\linewidth}}{
References: 1-- \citet{2010ApJ...710..432R} ; 2-- \citet{2003ApJ...583..451M} ; 3-- \citet{2008ApJ...684.1390R} ; 4-- \citet{2009ApJ...704..975J} ; 5-- \citet{2010ApJS..186...63R} ; 6-- \citet{2010ApJ...723..684B}.
}
\end{tabular}
\label{mytable:target_details}
\end{table}

\begin{figure}
\includegraphics[width=0.48\textwidth]{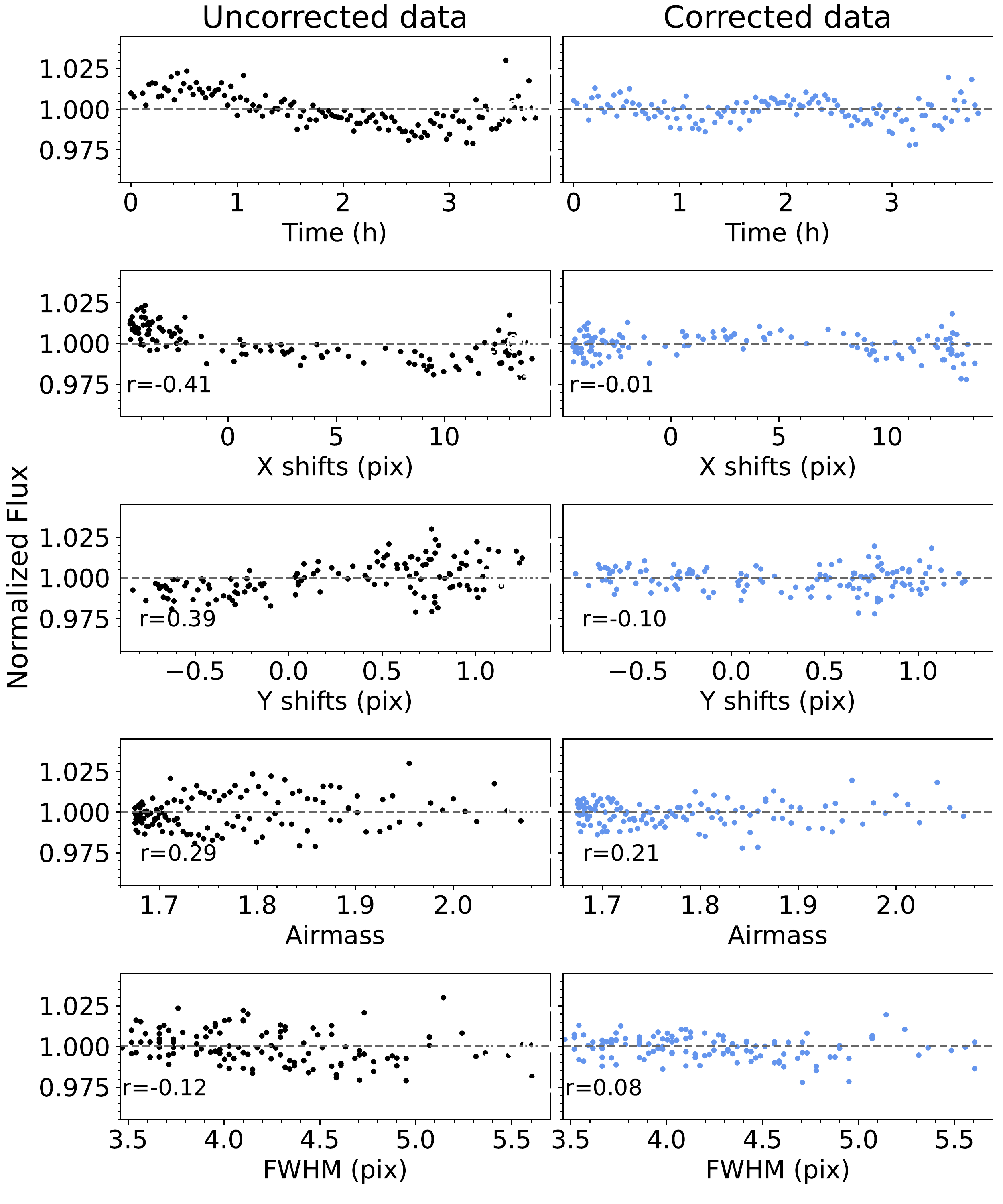}
\caption{Photometric light curve of 2MASS J14112131-2119503 from the WIYN 0.9 m telescope (top panels). The differential photometry in the left panels exhibits strong linear correlations between the normalized flux and the x and y pixel positions of the target on the detector (left, second and third panels from top). We de-trend the normalized flux for any target that shows an $| $r$|$ $\gtrsim$ 0.4 correlation with any of x or y pixel position, airmass, or PSF FWHM (Section~\ref{sec:anl-light-curve-extraction}).
The resulting corrected data (panels on the right) do not show any variability when these systematic effects have been resolved.}
\label{myfig:system}
\end{figure}

\section{Target Sample}
\label{sec:sample}

We aimed to monitor all $\geq$M7 ultra-cool dwarfs with measured projected rotation velocities of $v\sin i>30$\,km\,s$^{-1}$ and without reported rotation periods. We limited our sample to dwarfs brighter than 17.5 mag in the $I$ band, which generally allowed us to attain $<$1.5\% precision in a few minutes on 1 m-class telescopes by means of differential photometry. Thus, our survey sample comprised 13 ultra-cool dwarfs: nine M7--M9.5 dwarfs and four L0.5--L1.5 dwarfs, that are listed in Table \ref{mytable:target_details}.

Ultra-cool dwarfs with ages greater than 500 Myr are predicted to have radii of about 1\,$R_{\texttt{Jup}}$ \citep{2000ARA&A..38..337C}, which combined with $v\sin{i}$ $\geq 30$ km\,s$^{-1}$ constrain the rotational periods to $\lesssim4$ hr. Thus, our selection allows us to detect rotation-modulated variability within a single night of observations. 

\textit{Ages.} We checked whether any of our targets are known or suspected to be young based on probable membership in nearby young stellar moving groups, or on signatures of low surface gravity. Using the BANYAN $\Sigma$ tool \citep{2018ApJ...856...23G}, we found that most of our targets are not members of young associations, with the exception of 2MASS J16192988--2440469, which is an Upper Sco member \citep[$11\pm2$ Myr;][]{2017ApJ...838...73M}. In addition, 2MASS J14112131--2119503 was reported as an intermediate-gravity dwarf by \citet{2016ApJ...833...96L} and has a radius estimation of $1.98\pm0.44\,R_{\rm Jup}$ \citep{2015ApJ...810..158F}: larger than the $\approx1\,R_{\rm Jup}$ radius of $>$500~Myr-old higher-gravity ultra-cool dwarfs. We assume that the remaining 11 rapidly rotating targets in our sample are $>$500~Myr-old field ultra-cool dwarfs.

\textit{Binarity.} Our sample contains several binary or multiple systems. LP213-67 and LP213-68 were identified by \citet{2000MNRAS.311..385G} as a common proper motion pair separated by about 14.4\arcsec{}. Using adaptive optics observations, \citet{2003ApJ...587..407C} further resolved LP213-68 into an M8+L0 binary with the components separated by 0.12\arcsec{}. \citet{2017ApJS..231...15D} found that the components of LP213-68AB have an orbital period of nearly 6.6 years and masses of $97^{+6}_{-7} M_{\rm Jup}$ and $80\pm6 M_{\rm Jup}$. We estimate a magnitude difference of 1.2 mag in the $I$-band for the components of LP213-68AB by means of theoretical spectra. Additionally, \citet{2017AJ....154..151B} report strong perturbations in the 8-years astrometric data for 2MASS J23515044$-$2537367, which suggests the presence of a stellar companion with an orbital period longer than a decade. \cite{2006AJ....132..891R} observed this object using {\sl HST/NICMOS}, but did not resolve any companion. At the time of this work, the physical properties of the unresolved companion to 2MASS J23515044-2537367 are unknown. Finally, two of our targets---2MASS J13365044+4751321 and 2MASS J17054834-0516462 (hereafter J17054834-0516462, 2MASS is omitted in general)---have been observed at high angular resolution, but have not shown any companions \citep{2005ApJ...621.1023S,2006AJ....132..891R}. We are not aware of any surveys that have found companions to any of our other targets.

\section{Observations and Data Reduction}
\label{sec:obs}

\subsection{Ground-based observations}
Five of our targets were monitored with the Half Degree Imager (HDI) mounted on the WIYN 0.9 m telescope at Kitt Peak observatory (USA), and the other eight targets with the ANDICAM optical imager on the SMARTS 1.3 m telescope at Cerro Tololo Inter-American observatory (Chile). Observations at Kitt Peak were carried out in visitor mode from February 22$^{\rm nd}$--25$^{\rm th}$ 2018, while observations at Cerro Tololo consisted of 65 hours in service mode during semesters 2018A and 2018B. The observing log of our campaigns is shown in Table \ref{mytable:obs_data}.

HDI has a plate scale of 0.419\arcsec{}\,pixel$^{-1}$ and a field size of 29\minute$\times$29\minute. Observations were carried out using a Harris-$i$ filter, which has an effective wavelength of 7999 \AA\, with a pass band of 1368 \AA. Typical observations of a target covered 4-6 hours of continuous imaging. We used individual exposure times of 60-180 seconds, depending on the target's brightness. Typical seeing conditions were within 1.5\arcsec{}--2.5\arcsec{}. The sky conditions were mostly clear with very light cirrus cloud cover. 

ANDICAM has a plate scale of 0.371\arcsec{}\,pixel$^{-1}$ and a field of view of 6\minute$\times$6\minute. Observations were taken using a KPNO $I$ filter, which has an effective wavelength of 8151 \AA\, with a pass band of 1825 \AA. Each target was monitored over $\sim$2.5 hours. Typical seeing conditions were 1.0\arcsec-1.9\arcsec{}.

The optical SMARTS 1.3 m data has been bias-corrected and flat-fielded through their observations pipeline \footnote{http://www.astro.yale.edu/smarts/1.3m.html}. We used standard recipes for bias subtraction and flat field correction using the Image Reduction and Analysis Facility (IRAF) on the WIYN 0.9 m data. Every image was also inspected for defects. Images with significant background or detector noise, e.g., images from the very start or end of a night with $\geq$2 times lower signal-to-noise ratio (SNR) because of sky brightening, or images with defects such as weak horizontal striping, were removed from the data analysis. At most 10 per cent of the images (3--5 per observing sequence) were thus removed for poor quality. 

\subsection{Space-based observations}
Six of our targets (J00242463-0158201, LP213-67, LP213-68AB, J13365044+4751321, LP859-1, and J23515044-2537367; Table \ref{mytable:obs_data}) were observed by TESS with a cadence of 2 minutes as part of different guest observer programs\footnote{https://heasarc.gsfc.nasa.gov/docs/tess/approved-programs.html}. In addition, J14112131-2119503 is available in the 30-minutes Full-Frame Images (FFI) delivered by TESS. None of the remaining targets will be imaged by TESS at least until Sector 69.

The TESS data are released both in raw and pipeline calibrated data forms. For our purposes, the standard TESS pipeline\footnote{https://heasarc.gsfc.nasa.gov/docs/tess/documentation.html} is sufficient; only the calibrated Pre-search Data Conditioned Simple Aperture Photometry (PDCSAP) data were used in the following analysis. For J14112131-2119503 we used \texttt{eleanor} \citep{2019PASP..131i4502F} to extract its light curve from the FFI using a method similar to the TESS pipeline.

\begin{figure}
\centering
\includegraphics[width=0.45\textwidth]{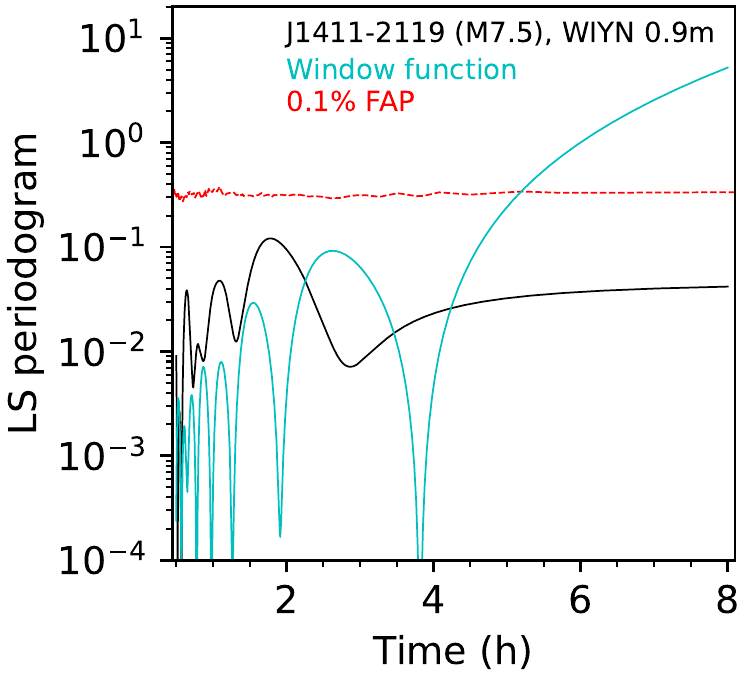}
\caption{LS periodogram (black) for the corrected light curve of J14112131-2119503 (Fig. \ref{myfig:system}). This target shows no significant periodicity in the ground-based WIYN 0.9 m data, as the periodogram power never exceeds the 0.1\% FAP level (red dashed line), and mimics the power distribution of the window function (cyan line).
}
\label{fig2}
\end{figure}

\begin{table*}
    \centering
    \caption{Observing log.}
    \begin{tabular}{cccccccccc}
    \hline 
         &Object & SpT & Observing & Filter$\rm ^a$  & Observation & N$_\texttt{c}$$\rm ^b$ & $\sigma_{\rm target}$$\rm ^c$  &$\sigma_{\rm err}$$\rm ^c$  & Airmass \\
         & &  & Date & & Time (h) & & (mmag) & (mmag) & \\
         \hline
         & 2MASS J00242463-0158201 & M9.5 &  11/08/18$\rm ^d$ & KPNO I & 3.66 & 3 &12 &10$\pm$4 & 1.14-1.68\\
         & &   &19/08/18$\rm ^d$ & KPNO I & 3.66 & 3  & 12 & 13$\pm$4 & 1.17-1.48 \\ 
         &  & & Sector 42 & TESS$\rm ^e$ & 551 & -- &-- &-- & ---\\
         &    & & Sector 43 & TESS$\rm ^e$ & 581 & -- &-- &-- & ---\\

         &LP213-67 & M7 & 22/02/18 &Harris i & 5.05 &6 & 7 & 4$\pm$1  &  1.02-1.28\\
         && &24/02/18 &Harris i & 5.09 &6 & 6 &5$\pm$1 & 1.01-1.47\\
         &   & & Sector 21 & TESS$\rm ^e$ & 648 & -- &-- &-- & ---\\
         
         &LP213-68AB & M8+L0 & 22/02/18 &Harris i & 5.05 &5 & 8 & 7$\pm$1 & 1.05-1.42 \\
         && &24/02/18 &Harris i & 5.09 &5 & 7 & 7$\pm$1 &  1.02-1.47 \\
         &    & & Sector 21 & TESS$\rm ^e$ & 648 & -- &-- &-- & ---\\
                  
         &2MASS J11593850+0057268 & L0 & 23/02/18 &Harris i & 4.98 &7 & 18 & 17$\pm$2 & 1.19-2.00\\
         
         &2MASS J13365044+4751321 & M7 & 25/02/18$\rm ^d$ &Harris i & 6.22 &7 & 6 & 5$\pm$1 & 1.10-1.49\\
         &  &   & Sector 16 & TESS$\rm ^e$ & 530 & -- &-- &-- & ---\\
         &  &   & Sector 22 & TESS$\rm ^e$ & 631 & -- &-- &-- & ---\\
         &  &   & Sector 23 & TESS$\rm ^e$ & 577 & -- &-- &-- & ---\\

         &2MASS J14112131-2119503 & M7.5 & 24/02/18$\rm ^d$ &Harris i & 3.87 & 5 & 11 & 7$\pm$1 & 2.01-2.14 \\
         && &10/05/18$\rm ^d$ &KPNO I & 3.25 & 5 & 7 & 8$\pm$3 &  1.04-1.18\\
         && &15/05/18$\rm ^d$ &KPNO I & 2.73 & 5 & 9 & 9$\pm$3 &  1.03-1.09\\
         &   & & Sector 11 & TESS$\rm ^e$ & 568 & -- &-- &-- & ---\\
 
         &2MASS J14310126-1953489 & M9 & 12/05/18 &KPNO I & 2.73 & 5 & 37 & 31$\pm$2 & 1.00-1.07 \\
         && &16/05/18 &KPNO I & 2.73 & 4 & 26 & 28$\pm$2 &  1.02-1.10\\
         
         &LP859-1 & M7 & 14/05/18 &KPNO I & 2.68 &4 & 22 & 21$\pm$1 & 1.01-1.10\\
         && &23/05/18 &KPNO I & 2.91 &4 & 7 & 7$\pm$3 & 1.01-1.13\\
         &    & & Sector 11 & TESS$\rm ^e$ & 575 & -- &-- &-- & ---\\
         &    & & Sector 38 & TESS$\rm ^e$ & 633 & -- &-- &-- & ---\\

         &2MASS J16192988-2440469 & M8 & 17/05/18 &KPNO I & 2.73 &4 & 86 & 84$\pm$3 & 1.01-1.31\\
         && &14/06/18 &KPNO I & 2.73 &4 & 46 & 47$\pm$7 & 1.00-1.08 \\
         
         &2MASS J17054834-0516462 & L1 & 24/05/18 &KPNO I & 2.73 &4 & 19 & 15$\pm$2 &  1.10-1.41\\
         && &26/05/18$\rm ^d$ &KPNO I & 2.73 &4 & 52 & 40$\pm$6 & 1.10-1.50\\
         
         &2MASS J20575409-0252302 & L1.5 & 20/06/18$\rm ^d$ &KPNO I & 2.73 &4 & 29 & 25$\pm$1 & 1.13-1.67\\
         && &22/06/18 &KPNO I & 2.73 &4 & 15 & 15$\pm$1 & 1.13-1.36\\
         
         &2MASS J23515044-2537367  & M9 & 25/06/18$\rm ^d$ &KPNO I & 2.73 &4 & 22 & 16$\pm$2 & 1.03-1.46\\
         && &24/07/18$\rm ^d$ &KPNO I & 2.73 &4 & 12 & 13$\pm$2 & 1.00-1.15\\
         &    & & Sector 2 & TESS$\rm ^e$ & 658 & -- &-- &-- & ---\\
         &    & & Sector 29 & TESS$\rm ^e$ & 572 & -- &-- &-- & ---\\

         \hline
    \end{tabular}
 \begin{minipage}{175.5mm}
Notes: 
$\rm ^a$The Harris filters are associated with the Kitt Peak WIYN 0.9m telescope while the KPNO I filters are associated with the SMARTS 1.3m telescope.\\
$\rm ^b$N$_\texttt{c}$ is the number of comparison stars used per field for the differential photometry described in Section \ref{sec:anl}.\\
$\rm ^c$$\sigma_{\rm target}$ indicates the scatter of the target light curve, while $\sigma_{\rm err}$ indicates the scatter of stars with flat light curves and brightness within $\pm$0.1 mags of that of the target. If were are no stars within $\pm$0.1 mags, $\sigma_{\rm err}$ indicates the scatter of the comparison star light curves.\\
$\rm ^d$Target lightcurve has been detrended for this observation date.\\
$\rm ^e$The analysis of the TESS data is described in Section \ref{sec:gp}.
\end{minipage}
    \label{mytable:obs_data}
\end{table*}

\section{Ground-based data analysis}
\label{sec:anl}

Our ground-based observations revealed periodicities in 3 of the 13 targets: details are given in this Section. Our subsequent analysis of the TESS light curves of seven of the sample targets significantly refined the periods and amplitudes for these three, and revealed three additional variables (Section~\ref{sec:spa}; Table~\ref{mytable:results}). That is, the TESS data supersede our ground-based data for the seven targets in common. We present our analysis of the ground-based data for completeness, but the reader is welcome to skip directly to Section~\ref{sec:spa} for all final period determinations. The ground-based data do allow us to correctly assign the TESS-detected variability to individual components of the $14\farcs4$ LP 213-67/68AB triple system, which is unresolved by TESS (Section~\ref{sec:interference}), and also include non-detections of variability in 6 of the sample targets not available from the TESS data releases. More generally, the ground-based data reveal the periods and amplitudes detectable with 1 m class telescopes for ultra-cool dwarfs.

\subsection{Light curve extraction}
\label{sec:anl-light-curve-extraction}
We obtained the light curves of our targets by means of differential photometry, i.e., the comparison of how stable the photometry of our target is in respect to that of a set of reference stars in the same field of view.  Following  \citet{2017MNRAS.472.2297M}, we identified all sources with SNR $>$ 5 in the field of view of each target. Then, we used the Astropy package {\tt Photutils} \citep{larry_bradley_2019_2533376} to perform circular aperture photometry on all our targets and sources in their fields. The aperture radius was chosen to maximize the SNR of the main target, which is then used for all other field stars. Typical radii values range between 3-6 pixels depending on target brightness, which roughly correspond to 1-1.5 times the typical full-width-at-half-maximum (FWHM) of the images. The photometry was sky-subtracted by computing the median value of the sky in an annulus with inner radius typically 12-20 pixels and a width of 8 pixels, depending on field density. Finally, the differential light curve of each target was constructed by dividing their fluxes by the sum of those of a set of comparison stars. These comparison stars were chosen among those stars slightly brighter than our targets, but far from the non-linear regime of the detector, that show the smallest standard deviation in their fluxes. Comparison star light curves were visually inspected for variability to ensure that any target variability detections are not spuriously created. We normalized the differential photometry of each target by dividing the fluxes by their average value. 

An example of these light curves is shown in the top panels of Figure \ref{myfig:system}, in which we also plot the differential photometry as a function of the pixel position in both the x- and y-axes, the airmass, and the FWHM to search for the presence of residual systematics that can mimic photometric variability. We used Pearson's r correlation coefficient to test for linear correlation attributable to any residual systematic \citep{taylor1990interpretation}. We found that only r-values $\gtrsim$ 0.4 can produce non-negligible variability in our data. For those cases, we determined the best fit straight line between the differential flux and the parameter involved in the correlation. Then, we detrended our data by dividing the differential photometry by the best fit. Figure \ref{myfig:system} shows an example of spurious photometric variability in one of our targets (panels on the left) and the detrending process described here (panels on the right). In total, the photometric data for 10 out of 23 campaigns were detrended. These are indicated in Table \ref{mytable:obs_data}, and their corresponding light curves are shown in Figure \ref{myfig:full_lc} of the Appendix \ref{app1}. Most of the targets with detrended light curves did not show real photometric variability, with the exception of J13365044+4751321, for which we corrected a linear trend with FWHM. We discuss this object further in Sections \ref{var} and \ref{sec:gp}.\\ 

\subsection{Search for photometric variability \label{var}}

We search for potential periodicities in our data by applying a Lomb-Scargle (LS) periodogram \citep{1976Ap&SS..39..447L,1982ApJ...263..835S} to all light curves shown in Figure \ref{myfig:full_lc}. For each LS periodogram we sample $10^4$ frequencies corresponding to periods between 0.5-8 hours, and also compute the window function \citep[e.g., ][]{2018ApJS..236...16V} and a 0.1\% False-Alarm-Probability (FAP), obtained from $10^4$ randomizations of the data \citep[e.g., ][]{2017MNRAS.472.2297M}.

\begin{centering}
\begin{figure}
\includegraphics[width=0.49\textwidth]{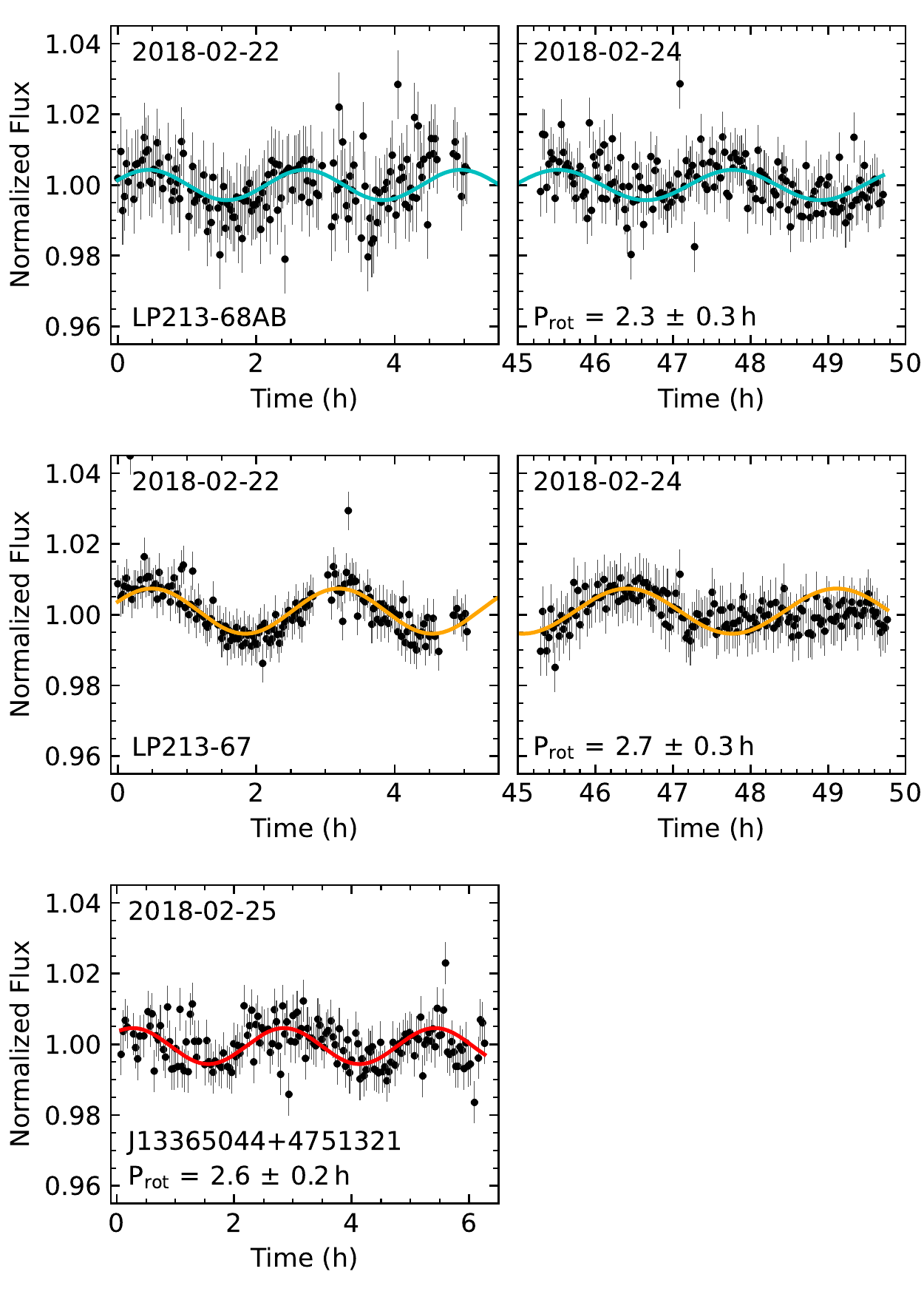}
\caption{Best fit sinusoid variations overlaid onto the calibrated data for LP213-68AB (top), LP213-67 (middle), and J13365044+4751321 (bottom). The best-fit rotation period for each target is indicated in the panels.}
\label{myfig:MCMC}
\end{figure}
\end{centering}

Three out of our 13 targets (LP 213-67, LP 213-68AB, and J13365044+4751321) showed significant peaks in the 2-3 h range, in agreement with expectations from their $v\sin{i}$ and radii. We fit a sine function (e.g., $y(t)=A\sin{\left(2\pi t/P +\phi\right)}+\mu$) to each light curve, and estimate variability properties by means of a $10^6$-step Markov Chain Monte Carlo (MCMC) process. We used flat priors for the amplitude (A), rotation period (P), phase ($\phi$), and mean ($\mu$), and allowed them to vary over the ranges: 0-1\%, 0.5-40 h, 0-2$\pi$, and 0.9-1.1, respectively. All MCMC fits for the variable targets and the corresponding rotation periods each night are shown in Figure \ref{myfig:MCMC}; the best-fit rotation periods are also listed in Table \ref{mytable:results}. We find amplitudes of variability in the range 0.4\%-0.8\%, and rotation periods of $2.7\pm0.3$ h, $2.3\pm0.3$ h, and $2.6\pm0.2$ h for LP 213-67, LP 213-68AB, and J13365044+4751321, respectively. For LP 213-67 and LP 213-68AB we used a single sine function to model the variability seen in both nights. While LP 213-68AB does exhibit the same variability amplitude on both nights (separated by $\sim$20 rotation cycles),  LP 213-67 shows some flattening in the observed variability from one epoch to the next one, which suggests the need for multi-epoch monitoring of these objects.

The remaining targets of our ground-based observations (J00242463-0158201, J11593850+0057268, J14112131-2119503, J14310126-1953489, LP859-1, J16192988-2440469, J17054834-0516462, J20575409-0252302, and J23515044-2537367) do not show any photometric variability larger than that seen in other field stars of similar brightness. This might be due to a lack of photospheric heterogeneities, or if present these may be too small to produce a measurable variability in our data sets.

\begin{centering}
\begin{figure*}
\includegraphics[width=0.9\textwidth]{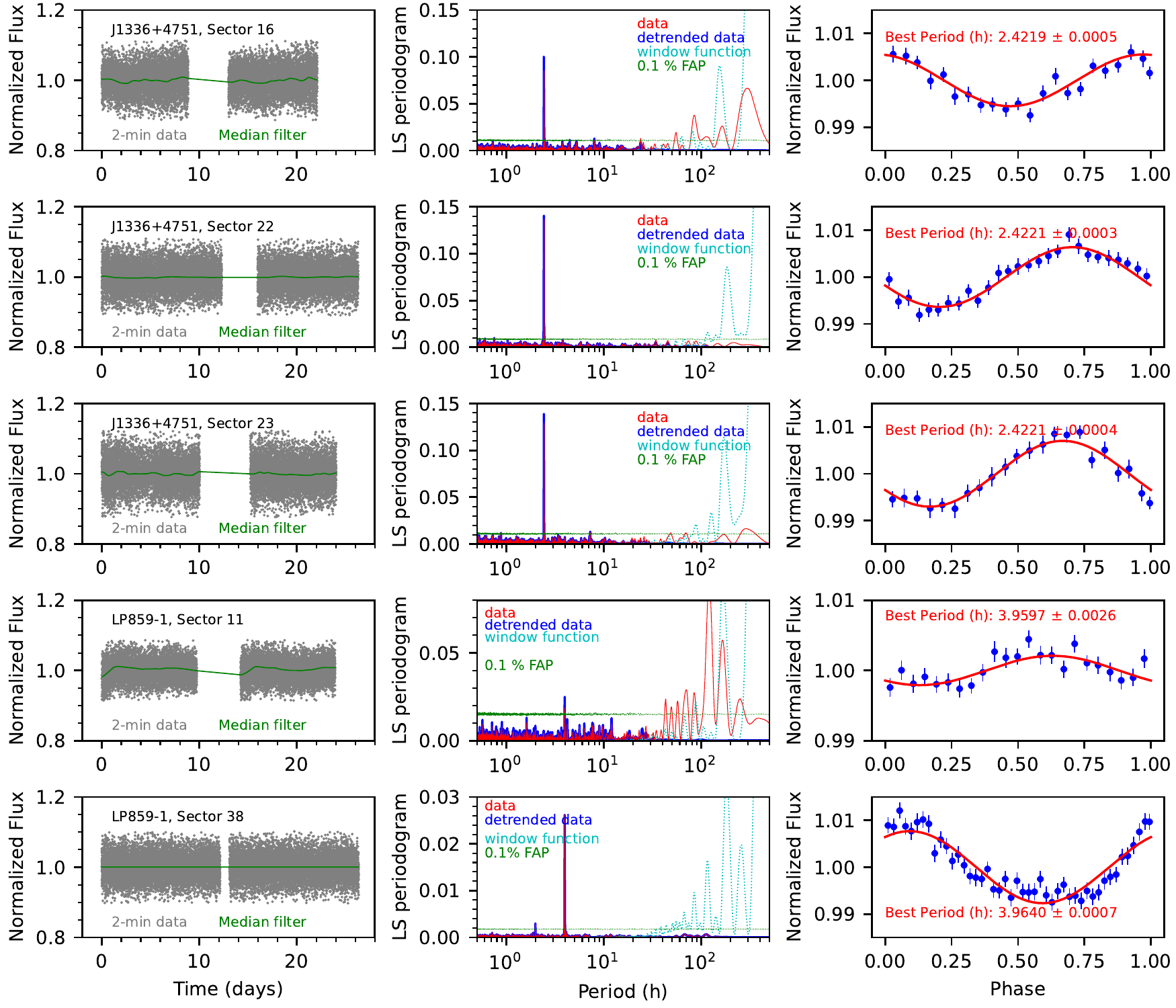}
\caption{{\it Left}: TESS data for two of the 13 targets in our ground-based sample. Each sector observation is shown in a different row, so some targets are shown on more than one row. Most of the light curves have a 2-minute cadence, except for J1411-2119 (Fig. \ref{fig:tess2}), which is at a 30-minute cadence. A 24-hour-long median filter is shown in green, which we use to trace potential residual systematics uncorrected by the pipeline. {\it Middle}: LS periodogram for the data before (red) and after (blue) removing the 1-day-long median filter. The window function and a 0.1\% FAP are also shown in cyan and green, respectively. {\it Right}: Phase-folded light curve using the rotation period derived with the gaussian process approach in Section \ref{sec:single}. The best fit sine curve is shown in red. A 300-point bin has been applied to the phase-folded data. For targets observed in multiple sectors, the light curves have been phase-folded using the same initial reference time.
}
\label{fig:tess1}
\end{figure*}
\end{centering}

\begin{centering}
\begin{figure*}
\includegraphics[width=0.9\textwidth]{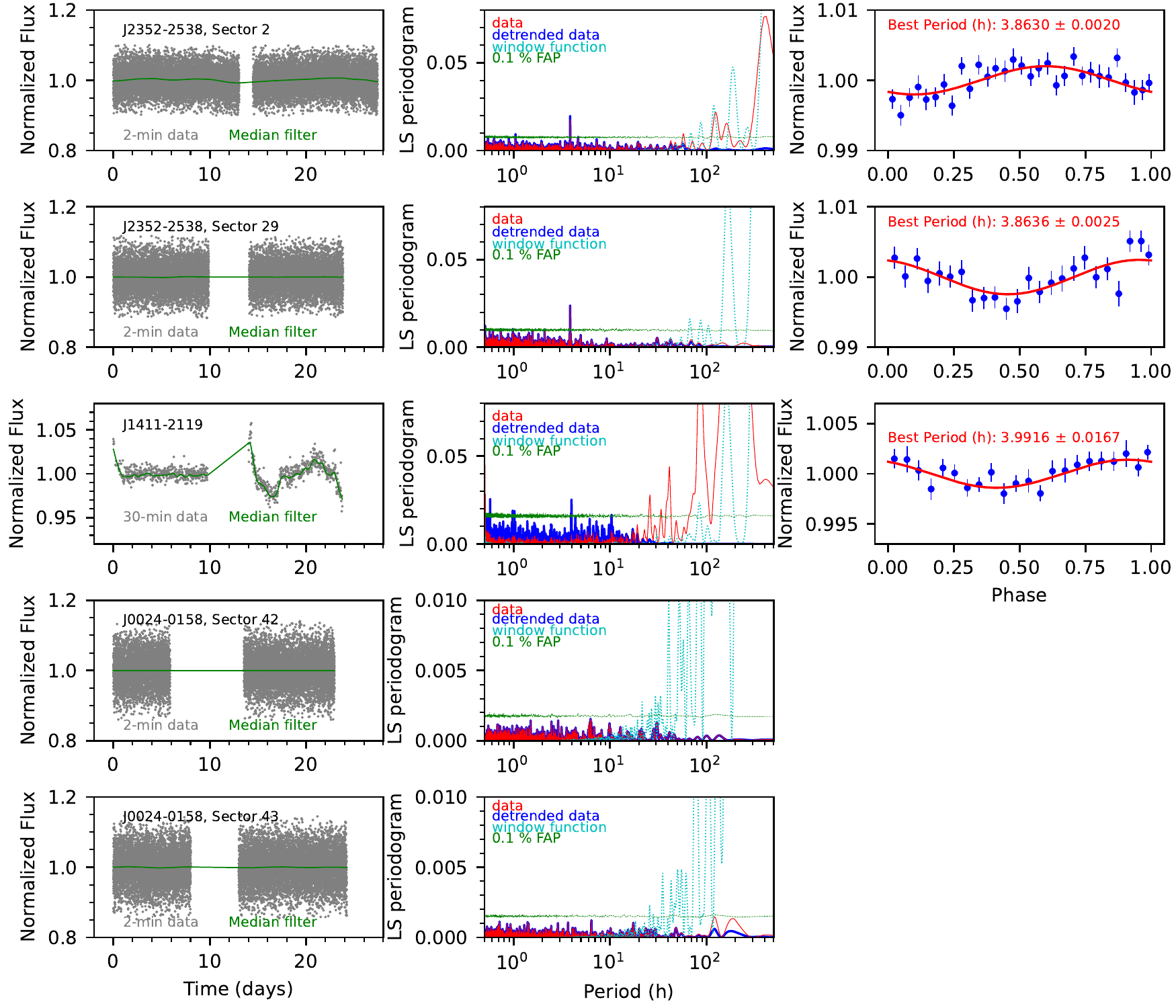}
\caption{Same as Figure \ref{fig:tess1}.}
\label{fig:tess2}
\end{figure*}
\end{centering}

\section{TESS data analysis}
\label{sec:spa}

The $\sim$27 days of continuous monitoring provided by TESS for each of its sectors has no parallel from the ground. For a typical rotation period of 4 h as expected for our rapidly rotating targets, TESS would observe 162 continuous rotation cycles in one single sector, allowing us to search for periodicities in the data with amplitudes of variability much smaller than those measurable from the ground \citep[e.g.,][]{2021A&A...651L...7M}. At the time of this work, 2-minute-cadence TESS data were available for J00242463-0158201, LP213-67 and LP213-68AB (spatially unresolved triple), J13365044+4751321, J23515044-2537367, and LP859-1 (see Table \ref{mytable:obs_data}). In our analysis we use the Pre-search Data Conditioned Simple Aperture Photometry (PDCSAP) light curves extracted via the TESS pipeline. We also found 30-minutes-cadence observations for J14112131-2119503 in the FFI delivered by TESS, and extracted the light curve using {\tt eleanor} \citep{2019PASP..131i4502F} in a similar way as done by the TESS pipeline. We removed any data points that the pipeline had flagged as having lower quality. This was the case for less than 1\% of the data for each target. Panels on the left of Figures \ref{fig:tess1} and \ref{fig:tess2} show the extracted light curves for J00242463-0158201, J13365044+4751321, J14112131-2119503, J23515044-2537367, and LP859-1. In the case of LP213-67 and LP213-68AB, their small (14.4\arcsec) separation places the combined light of the triple system in the same  21$\arcsec\,\times$\,21\arcsec pixel of TESS. We show the compound light curve of LP213-67 and LP213-68AB in the top panel of Figure \ref{lp1}, and discuss this system in more detail in Section~\ref{sec:interference}.

\subsection{Approximate light curve analysis with detrending and an LS periodogram}
\label{sec:ls}

We first conduct an approximate analysis of the light curves to test for periodicities in the TESS data. Targets that do display such periodicities are then analyzed with a more robust Gaussian processes-based approach (Section~\ref{sec:gp}).

\begin{centering}
\begin{figure}
\includegraphics[width=0.49\textwidth]{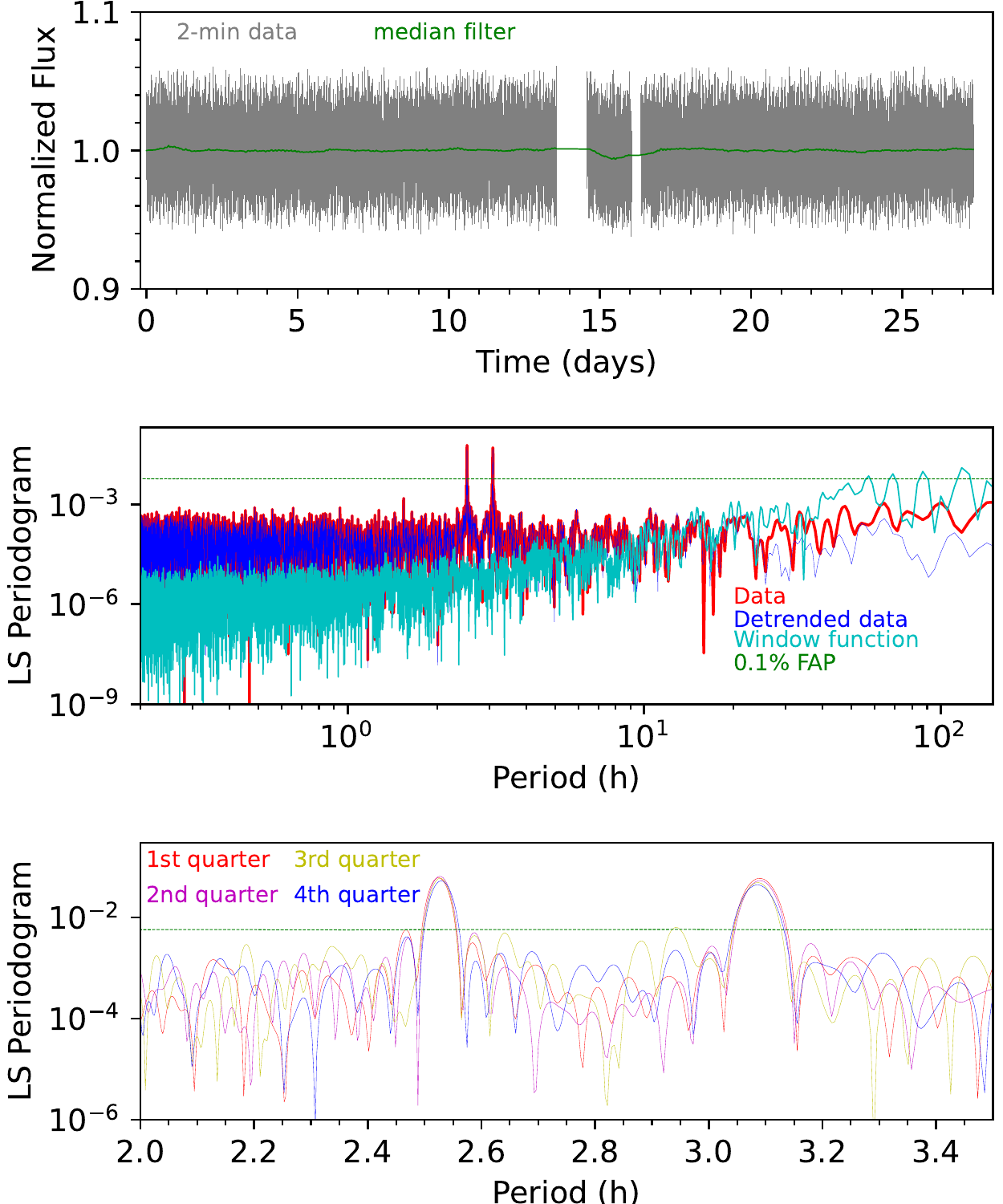}
\caption{{\it Top}: TESS light curve for the combined light of the triple system LP213-67 and LP213-68AB. Vertical lines denote individual uncertainties, and the green line is a 1-day-long median filter. {\it Middle}: LS periodogram for the TESS light curve shown in the top panel (red) and after detrending with a median filter (blue). The window function (cyan) and associated 0.1\% FAP (green) are also shown. {\it Bottom}: LS periodograms for each of the of four different quarters of the data, zoomed in on the significant peaks at $\sim$2.5 h and $\sim$3.1 h present in all the quarters of the data.}
\label{lp1}
\end{figure}
\end{centering}

The TESS light curves shown in Figures \ref{fig:tess1} (and \ref{fig:tess2}, left) and \ref{lp1} (top) exhibit some temporal structure on time scales of $>$1 day, which are likely associated with spacecraft momentum dumps, changes in the camera temperature, or Moon-Earth scattered light that is not fully removed by the pipeline \citep{2019ESS.....433312V}. We do a first-order correction of these trends by removing from the data a 24-h median filter, which is also shown in the figures (green lines). Then, we search for periodicity in the data by computing the LS periodogram of the raw and detrended data. These periodograms are plotted in the central panels of Figures \ref{fig:tess1} (and \ref{fig:tess2}) and \ref{lp1} together with their associated window functions and 0.1\% FAP, computed as in Section \ref{var}. 

We find significant periodicities in for J13365044+4751321, J14112131-2119503, LP859-1, J23515044-2537367 (central panels of Fig. \ref{fig:tess1} and \ref{fig:tess2}), and LP213-67 and LP213-68AB (central panel of Fig. \ref{lp1}), but no significant periodicity in any of the sectors of J00242463-0158201. In the case of J13365044+4751321, LP213-67 and LP213-68AB the periodicities fall in the same range of the periods measured in our ground-based data, while for J14112131-2119503, LP859-1, and J23515044-2537367 the detected periodicities were not seen in our campaign from the ground. In all the cases the significant periods are contained in the range of the expected rotation period for our targets. 

The first-order correction with a median filter is useful to quickly identify potential significant periodicity in the data by using a LS periodogram, especially in the case of J14112131-2119503 for which only the detrended data show a significant period in the LS periodogram. However, our choice of the 24-h median filter width is somewhat arbitrary. What is more, the filter could remove some real periodicity. Thus, we do a more sophisticated modelling of the variability and the noise in the data in Section \ref{sec:gp} by using Gaussian Processes regression as described in \citet{2017MNRAS.466.4250L} and as done for TESS data and ultra-cool dwarfs in \citet{2021A&A...651L...7M}.

\begin{centering}
\begin{figure*}
\includegraphics[width=0.7\textwidth]{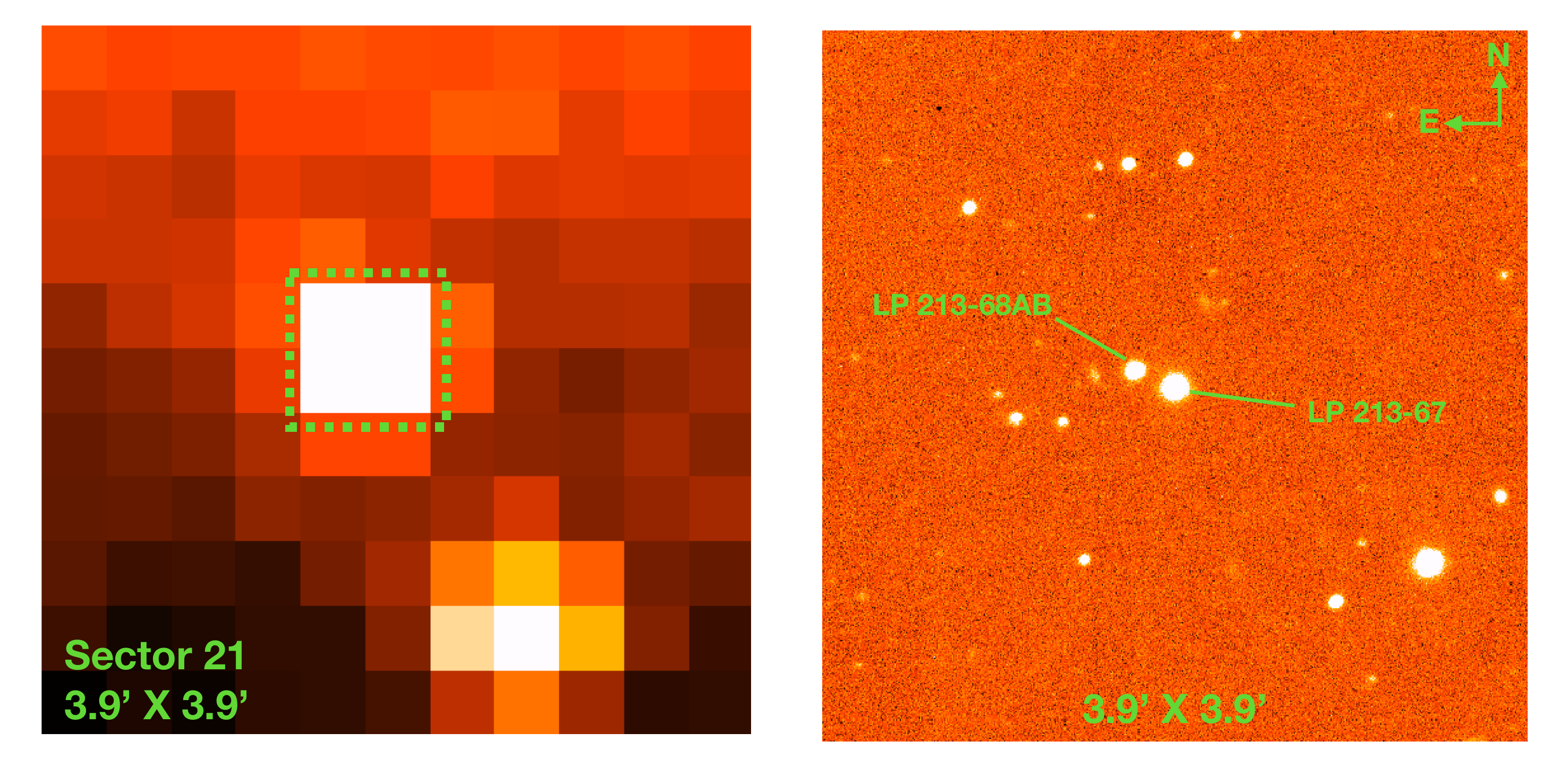}
\caption{Field of view of the system LP 213-67 and LP 213-68AB as seen from TESS (left; dashed green square) and from the ground (right; WIYN/HDI on 22nd February 2018). Both images are centered in the position of LP 213-67 and their fields of view have the same size ($3\farcm9\times3\farcm9$).}
\label{fov}
\end{figure*}
\end{centering}

\begin{centering}
\begin{figure*}
\includegraphics[width=0.85\textwidth]{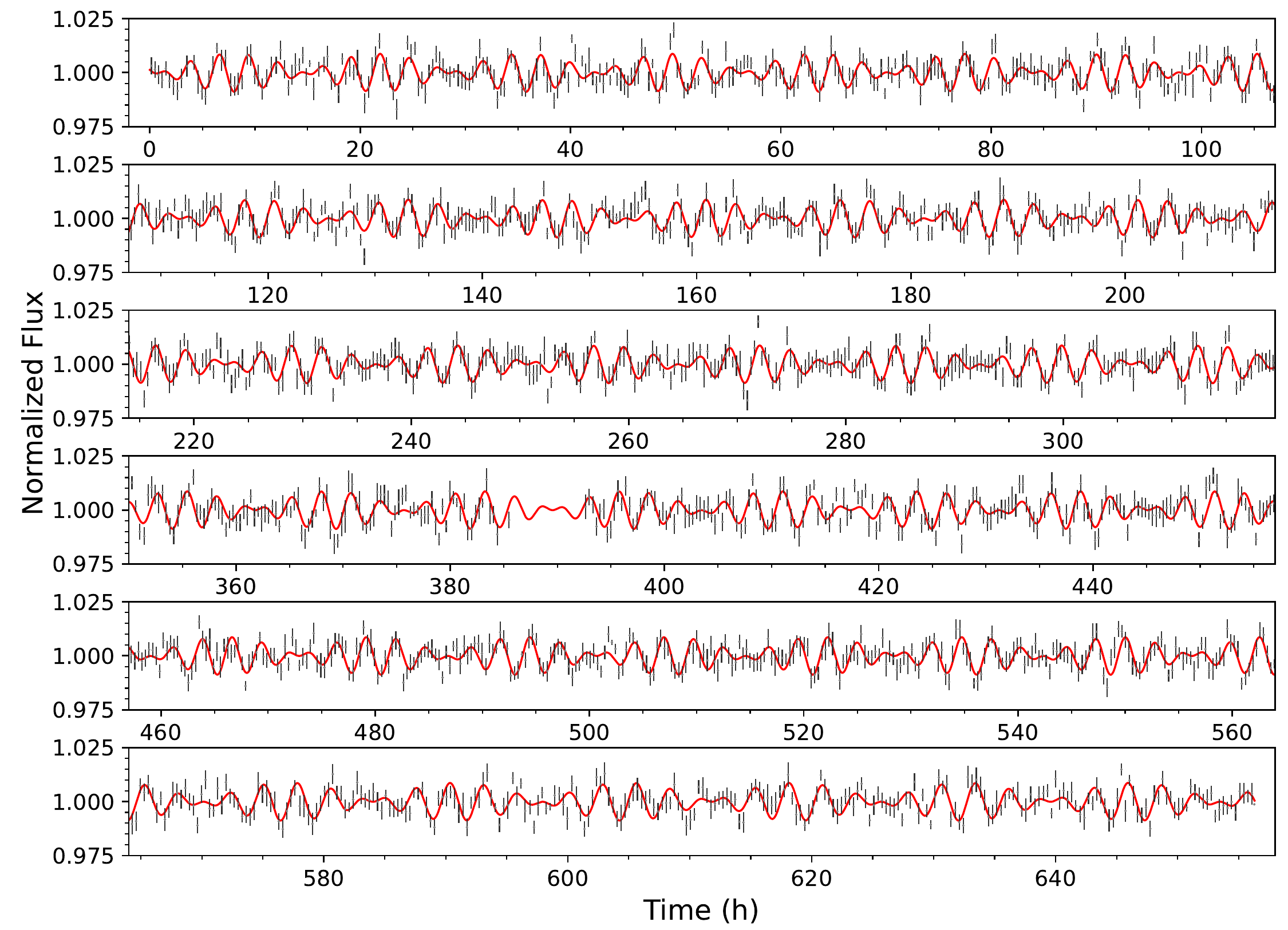}
\caption{TESS light curve for the combined flux of LP 213-67 and LP 213-68AB binned every 10 images for a resulting cadence of 20 min. The best fit for the two-period model from Section \ref{sec:interference} is  shown in red.}
\label{lp3}
\end{figure*}
\end{centering}

\subsection{Robust light curve modelling using Gaussian Processes regression \label{sec:gp}}

\subsubsection{Single targets \label{sec:single}}

Gaussian processes (GP) are a class of Bayesian non-parametric models that allow simultaneous capture of the expected behaviour of data in a time series and their associated noise due to systematics \citep{2012RSPTA.37110550R,2014sdmm.book.....I}. The part related to the general trend of the data is usually called the {\sl mean} function, while the part related to the noise is usually referred to as the {\sl kernel} function and deals with the covariance of the data. The  parameters associated with the kernel function are called hyperparameters. Following \citet{2021A&A...651L...7M}, we account for red noise in the TESS data, i.e., the low-frequency trends removed by the median filter in Figures \ref{fig:tess1}, \ref{fig:tess2} (left) and \ref{lp1} (top), by using a Mat\'ern-3/2 kernel \citep{2006gpml.book.....R} for each component ($k(t_{i}, t_{j})$) of the covariance matrix of the data:

\begin{equation}
k(t_{i}, t_{j}) = a^{2} \left( 1 + \sqrt{ \frac{3r^{2}}{\tau^{2}} } \right) \exp{\left( - \sqrt{\frac{3r^2}{\tau^2}} \right)},
\label{eq1}
\end{equation}

\noindent
where $r^{2} = (t_{i} - t_{j})^{2}$, $t_{i}$ is the time for data point $i$, and $a$ and $\tau$ are the typical amplitude and time scale of the red noise. We also add a white noise component ($\sigma$) that only contributes to the diagonal elements of the covariance matrix:

\begin{equation}
K_{ij} = \sigma^{2} \delta_{ij} + k(t_{i}, t_{j})
\label{eq:Kij}
\end{equation}

\noindent
Equation \ref{eq:Kij} thus defines the covariance matrix of our kernel function. For the mean function of our model we adopt a sum of two functions. The first one is a time-dependent sine function, as is appropriate for most optical rotation-modulated light curves for late-M and early-L dwarfs (e.g., see Figure \ref{myfig:MCMC} above or analyses in, e.g., \citealt{1997MNRAS.286L..17M,2013ApJ...779..101H,2017MNRAS.472.2297M}) The second one is a time-invariant $\mu$ parameter that describes a flat light curve. The mean function $f(t)$ is thus:

\begin{equation}
f(t) = \mu + A \sin{ \left[ 2\pi \left( \frac{t}{P} \right) + \phi \right] }
\label{eq:mean_function}
\end{equation}

We used the {\tt celerite} package  \citep{2017AJ....154..220F} to compute the GP (assuming flat and periodic light curves), and {\tt emcee} \citep{2010CAMCS...5...65G,2013PASP..125..306F} to run the MCMC process that fits our models to the data of each sector for J13365044+4751321 (3 sectors), J14112131-2119503, J23515044-2537367 (2 sectors), and LP859-1 (2 sectors). We analyze the combined light curve of LP213-67 and LP213-68AB separately in Section \ref{sec:interference}. For the MCMC we used 32 walkers with 500 iterations for the burn-in stage and 5000 iterations for the full process. We used log-uniform priors with values between 0.01 and 10 times the standard deviation of the light curve for the hyperparameter $a$, 2 minutes and 27 days for $\tau$, and a broad range of ($10^{-20}$,10)  for $\sigma$, while for the parameters in eq. \ref{eq:mean_function} we adopted the following ranges: 0.9-1.1 ($\mu$), $10^{-5}$-$10^{-1}$ ($A$), 0-2$\pi$ ($\phi$), and 0.5-40 h for $P$ \citep[as usually seen in the field for other ultra-cool dwarfs, e.g.,][]{2021AJ....161..224T}. 

We investigated whether a flat (Model 1) or periodic (Model 2) light curve is more favoured by the data in each sector by evaluating the Bayesian Information Criterion \citep[BIC,][]{1978AnSta...6..461S} for each model. In general, $\Delta{\rm BIC}={\rm BIC}_{\rm Model\,1}-{\rm BIC}_{\rm Model\,2}<2$ indicates no significant preference of the data for either model, $2<\Delta{\rm BIC}<6$ suggests a preference for Model 2, $6<\Delta{\rm BIC}<10$ points to further increasing support for Model 2, and $\Delta{\rm BIC}>10$ indicates that Model 2 is strongly favoured by the data. 

The MCMC processes quickly converged to a solution for J13365044+4751321 (3 sectors), J14112131-2119503, J23515044-253736 (2 sectors), and LP859-1 (2 sectors), and their $\Delta$BIC strongly favours the periodic models. The results from the GP fitting and MCMC parameter estimation processes for all unresolved TESS targets are included in Table~\ref{mytable:results}.
We find rotation periods very similar to those shown in the periodograms of Figures \ref{fig:tess1} and \ref{fig:tess2} for all of these targets. Moreover, the periods retrieved for targets with data in multiple sectors are identical down to $\sim10^{-3}$ h, which supports the evidence for real periodicies in the targets. Once we determined best-fit noise and variability models, we subtracted from our data the best-fit noise model, and phase-folded the corrected data on the best-fit rotation period (for objects with several sectors, we used the weighted average to compute the final rotation period). We show these light curves in the right-hand panels on of Figures \ref{fig:tess1} and \ref{fig:tess2}. 

The data shown in Figure \ref{fig:tess1} (right) for the three sectors of J13365044+4751321 were phase-folded using the beginning of sector 16 as the reference time ``zero'', which allow us to investigate the stability of the photometric variability over multiple months. The photometric variability over the consecutive sectors 22 and 23 stays in phase, which indicates that its origin is likely stable on time scales of more than a month (a few hundred rotation cycles). On the other hand, observations for sector 16 started $\sim$162 days earlier than those for sector 22, and the phase-folded data for J13365044+4751321 in sector 16 show a phase offset of $0.28\pm0.04$ compared to the phase-folded data in sectors 22 and 23. At a first glance, this offset might suggest that the light curve is not stable on time scales longer than six months. However, between the beginning of observations in sectors 16 and 22 there are nearly 1600 rotation cycles. When combined with the 0.0003 h uncertainty on the period, we obtain an uncertainty on the phase-folding of the data: 1600$\times$0.0003 = 0.48 h or 0.20 in phase. This is comparable to the observed phase offset in the light curves of J13365044+4751321 between sectors 16 and 22-23. Hence, it is likely that this light curve is stable on times scales of years, similarly to other fast rotators, such as the M8.5 dwarf TVLM 513-46 \citep{2014ApJ...788...23W} or the L1 dwarf WISEP J190648.47+401106.8 \citep{2013ApJ...779..172G}.

We can repeat the previous analysis for LP859-1 and J23515044-253736 that have data in two sectors separated by about two years. However, despite the fact that their period uncertainties are only a few seconds, the propagated phase uncertainties over nearly two years are too large to assess whether the light curves remain in phase or not. It is still remarkable that the retrieved rotation periods for LP859-1 and J23515044-253736 are very similar at each of the respective observing epochs, and also in good agreement with the expected rotation period from their observed $v$\,sin$i$ and radii. Nonetheless, the variability amplitude of LP859-1 is significantly different between the two epochs two years apart, which could be evidence for time-dependence of the spot pattern on the stellar surface.
 
\subsubsection{LP 213-67 and LP 213-68AB: spatially unresolved ultra-cool dwarfs produce a beat pattern in the TESS light curve
\label{sec:interference}}

LP 213-67 (M7) and LP 213-68AB (M8+L0) form a triple system of ultra-cool dwarfs. The M7 component and the (unresolved) M8+L0 pair are separated by only 14\farcs4, meaning that light from the entire system is contained in a single pixel of TESS as seen in Figure \ref{fov}. Because of this peculiarity we discuss this system separately. 

The 2-min TESS light curve of the combined fluxes of this system is shown in the top panel of Figure \ref{lp1}. The light curve also exhibits some trends on time scales of a day (green line), similarly to the TESS light curves of the other single targets (Fig.~\ref{fig:tess1}, \ref{fig:tess2}). The middle panel of Figure \ref{lp1} shows the LS periodogram for the raw (red) and detrended (see Section~\ref{sec:ls}; blue) light curve as well as the associated window function (cyan), and 0.1\% FAP level (green dashed line). This Figure shows two significant peaks at $\approx$2.5 h and $\approx$3.1 h, which are close to the rotation periods detected in our ground-based data for LP 213-68AB (2.3 h) and LP 213-67 (2.7 h). We check that these periodicities are present in all the TESS data by splitting this into four quarters and computing their associated LS periodograms, which we show in the bottom panel of Figure \ref{lp1}. 

A closer look into this data set (Figure \ref{lp3}) reveals that some epochs show a clear quasi-sinusoidal modulation, while other epochs show negligible modulation. \citet{2021A&A...651L...7M} reported a similar behaviour for the M7.5 binary VHS J1256-1257AB and showed that the light curve can be explained by the interference of two waves of the same amplitude and similar frequencies. We adopt this approach to model the combined light curve of LP 213-67 and LP 213-68AB, which are comparable in brightness. Thus, we modify the GP presented in section \ref{sec:gp} by using two sines for the mean function in eq. \ref{eq:mean_function}, with independent amplitudes, rotation periods and phases:

\begin{equation}
f(t) = \mu + A_1 \sin{ \left[ 2\pi \left( \frac{t}{P_{\rm rot,1}} \right) + \phi_{\rm rot,1} \right] } + A_2 \sin{ \left[ 2\pi \left( \frac{t}{P_{\rm rot,2}} \right) + \phi_{\rm rot,2} \right] }
\label{eq3}
\end{equation}

Similarly to section \ref{sec:single}, we used a Mat\'ern-3/2 kernel (eq. \ref{eq1}) to model the systematics in the TESS light curve and run an MCMC. Unsurprisingly the best solution  contains the  two rotation periods identified in the LS periodogram of Fig. \ref{lp1}: $2.5267\pm0.0002$ h and $3.0889\pm0.0002$ h. Interestingly, both amplitudes of variability converge to the same value: 0.44$\pm$0.01\%. There is no astrophysical reason to assume that both targets will have the same amplitude of variability. Thus, this result likely reflects the fact that TESS provides enough sensitivity to discern periodicities, but the photometry is not precise enough to distinguish different amplitudes of variability in our (faint) targets. The TESS light curve does not allow us to assign which of the periods corresponds to each component of the triple system. However, we use the results from our ground-based photometric analysis (Fig. \ref{myfig:MCMC}) to assign $2.5267\pm0.0002$ h to LP 213-68A and $3.0889\pm0.0002$ h to LP 213-67. In Figure \ref{lp3} we also plot the best fit for the two-period model (eq. \ref{eq3}).

\begin{centering}
\begin{figure*}
\includegraphics[width=0.95\textwidth]{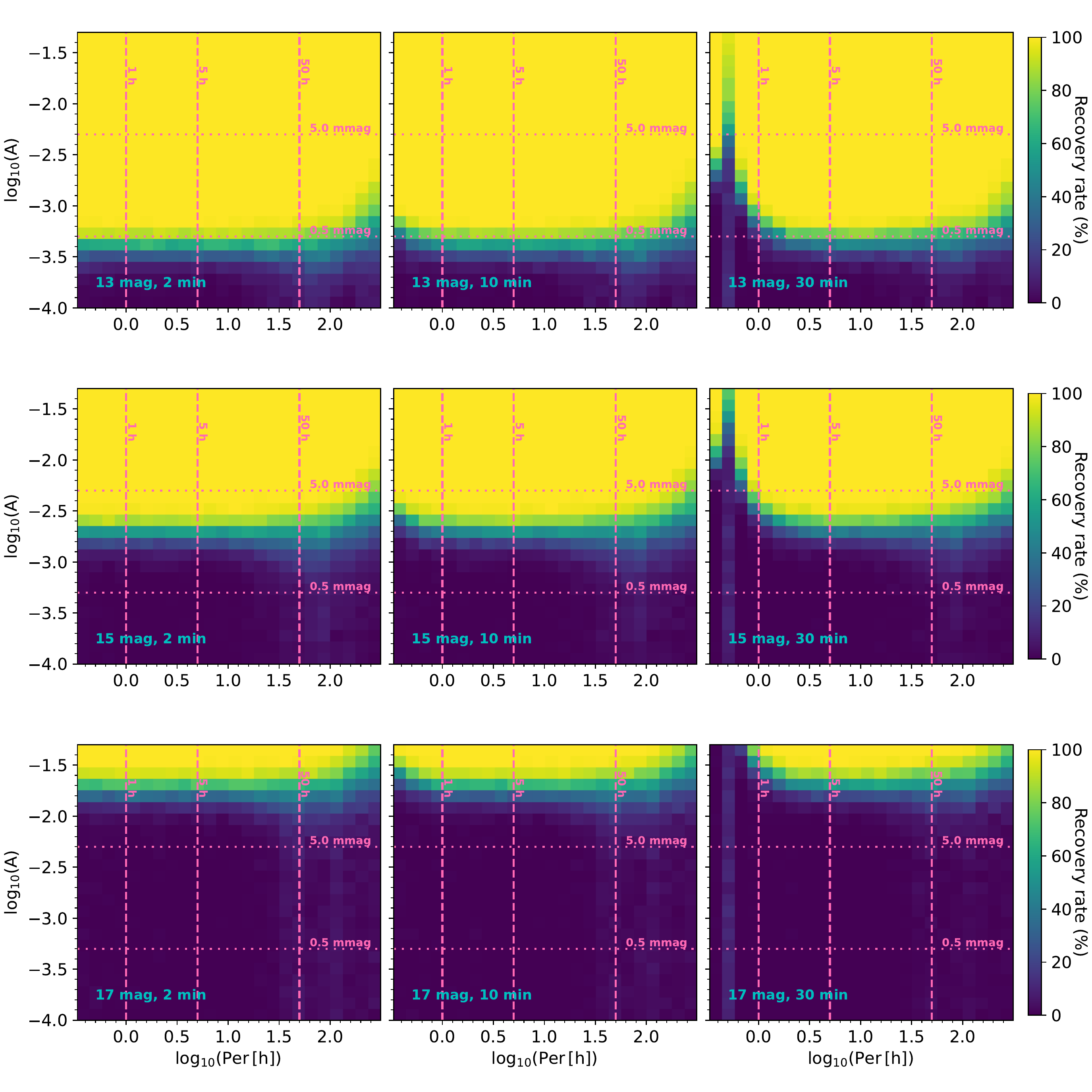}
\caption{Sensitivity maps for sinusoidal variability of amplitude A ($y$ axis) and period P ($x$ axis) from 200,000 simulated TESS light curves for different TESS magnitudes (top to bottom) and cadences (left to right). The amplitude is determined as a fraction of the mean flux level. The recovery rate is colour-coded as in the legends on the right.  An object's variation parameters are considered recovered if they are within 5\% of their input values for the simulation. We set a minimum recovery rate of 95\% at any of the simulated periods as a requirement for periodicity to be detected.}
\label{maps}
\end{figure*}
\end{centering}

\begin{centering}
\begin{figure}
\includegraphics[width=0.49\textwidth]{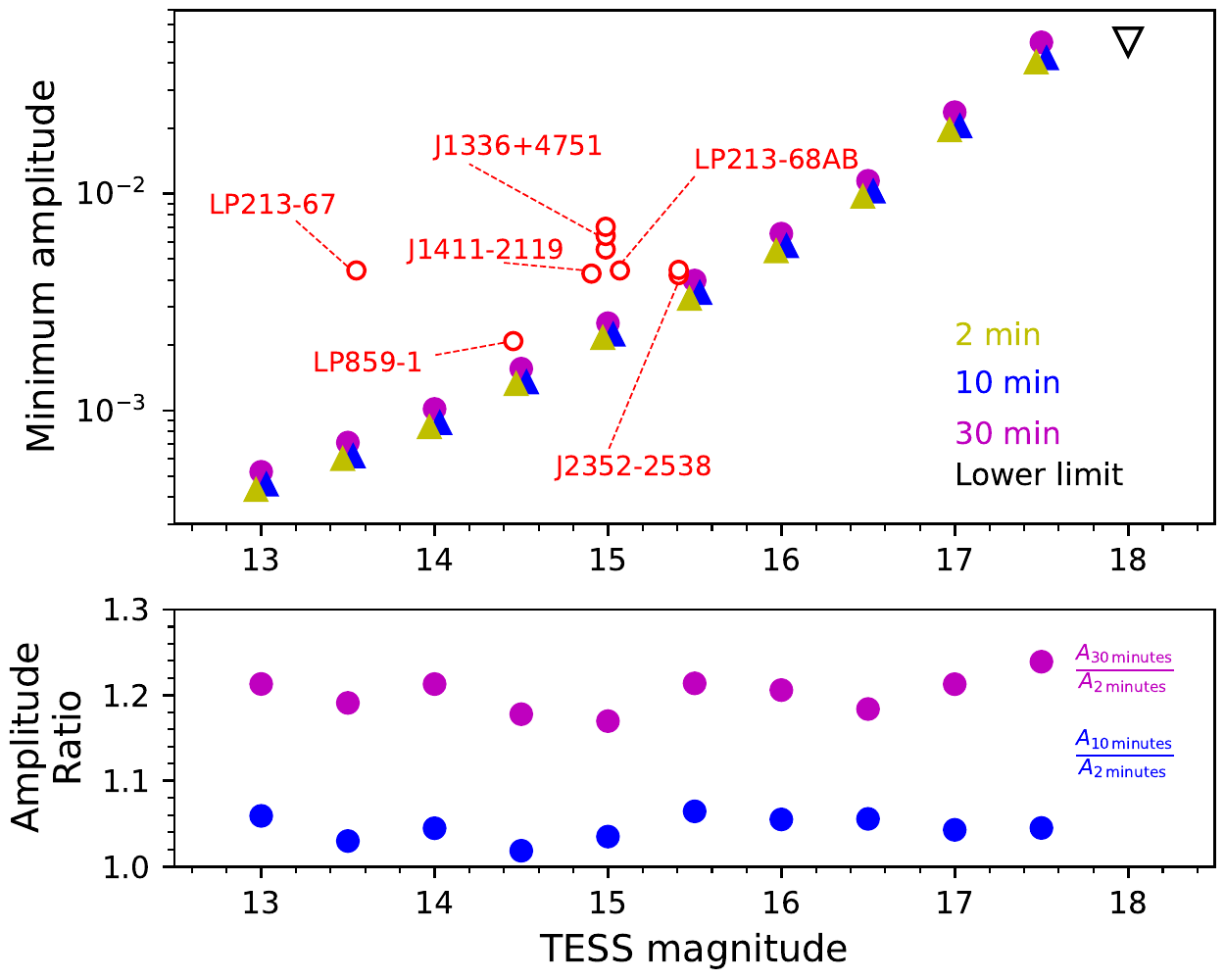}
\caption{{\it Top}: Minimum photometric amplitude for sinusoidal variability detectable by TESS (with a 95\% recovery rate) for a cadence of 2 min, 10 min, and 30 min as a function of stellar magnitude from the amplitude-period sensitivity maps in Figure \ref{maps}. As in Figure~\ref{maps}, the amplitude is determined as a fraction of the mean flux level.
{\it Bottom}: Ratio between the minimum amplitude of variability detected in light curves with a cadence of 10 min and 2 min (blue) and 30 min and 2 min (magenta). The most recent TESS FFI with $\sim$10-min cadence allow for ($>$0.5-hour period) variability searches in ultra-cool dwarfs comparable to those with 2-min light curves.}
\label{smr}
\end{figure}
\end{centering}

\begin{centering}
\begin{figure}
\includegraphics[width=0.49\textwidth]{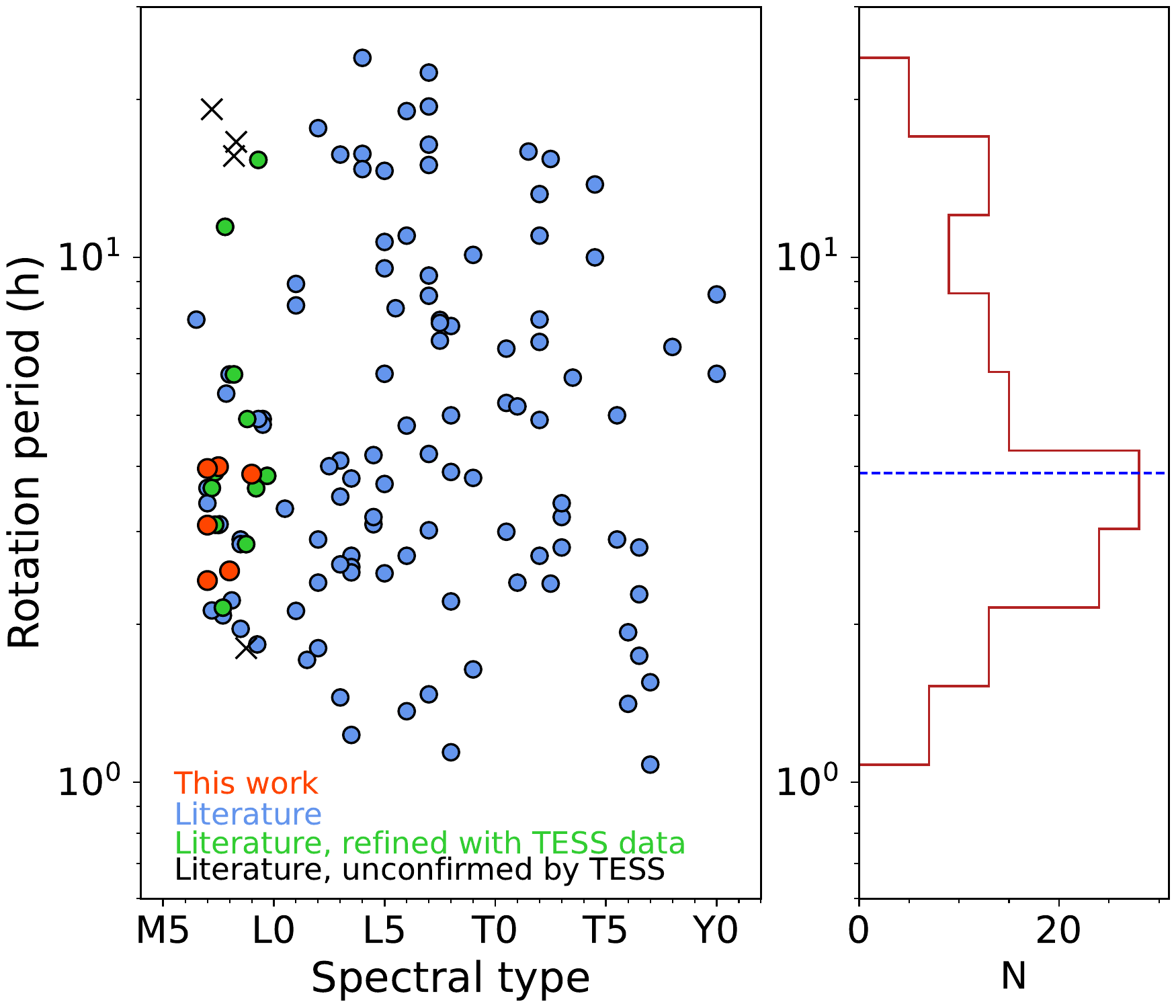}
\caption{
{\it Left}: Rotation period as a function of spectral type for all 128 periodically variable M7–Y0 dwarfs known as of this writing. Data are taken from compilations listed in \citet{2017ApJ...834...85N,2017MNRAS.472.2297M,2017ApJ...840...83M, 2021AJ....161..224T} and the new detections reported in \citet{2021A&A...651L...7M,2022ApJ...924...68V,2022MNRAS.tmp..993A}. New photometrically variable targets reported in this work are indicated by red circles, while refined rotation periods for known variable ultra-cool dwarfs are indicated with green circles. We use crosses for targets reported to be photometrically variable in the literature that we could not confirm with TESS data. {\it Right}: Distribution of all known photometric periods of field M7-Y0 dwarfs. The median rotation period for field ultra-cool dwarfs is 3.9 hr (blue dashed line).}

\label{fig:rot}
\end{figure}
\end{centering}

\section{Discussion}
\label{disc}
Sections \ref{sec:single} and \ref{sec:interference} show the synergy between TESS, which is able to unveil small periodicities that would be missed from the ground (e.g., J14112131-2119503, LP859-1, or J23515044-253736), and ground-based monitoring to  safely discern the source of variability when several objects fall inside the same pixel of TESS (e.g., LP 213-67 and LP 213-68AB). Given that TESS, with its throughput peaking near 900~nm, is optimized to observe red stars, it can allow us to explore the distribution of rotation periods of the brightest ultra-cool dwarfs. 

\subsection{Sensitivity of TESS to ultra-cool dwarf photometric periodicities and amplitudes}

To explore the range of amplitudes and rotation periods that can be detected by TESS we perform Monte Carlo simulations. First, we simulate a 27 day-long TESS time series (e.g., like the data shown in Figs. \ref{fig:tess1}, \ref{fig:tess2}, left) with a cadence of 2-min and a gap of 3 days in the middle of the light curve to simulate the time in which data are sent to the ground. From each of the two halves of the resulting time series we further remove some contiguous data to simulate bad quality data, which are removed by the pipeline. These additional gaps are randomly located in each half of the light curve, with a duration randomly sampled between 5--72 h. Second, we inject in our time series the expected white noise from the TESS model presented in \citet{2018AJ....156..102S} and \citet{2017zndo....888217B} for TESS magnitudes from 13 to 19 in steps of 0.5 mag. Third, we add red noise by using the Mat\'ern-3/2 kernel of eq. \ref{eq1}. We randomly sample the kernel hyper-parameters from a log-uniform distribution with limits  0.5--5 times the expected white noise for $a$ and 0.5--13 days for $\tau$, which matches the typical values found in our analysis in Section \ref{sec:gp}. Finally, we inject a sine wave with mean value of 1 and amplitude, rotation period and zero phase randomly sampled from [0.01--5]\% (log-uniform distribution),  [0.3--312] h (log-uniform distribution), and [0--2$\pi$] (uniform distribution). 

We simulated 200,000 light curves at each half-magnitude step between 13.0 and 19.0. In addition to simulating 2-min cadence, we also binned the light curves every 10 and 30 min to simulate the FFI delivered by TESS. Fitting a GP to these light curves would be extremely expensive in terms of computing time. Instead we repeated our approximate analysis from Section \ref{sec:spa} (left and middle panels of Figures \ref{fig:tess1} and \ref{fig:tess2}). Specifically, we: i) removed a 1-day long median filter to the data, ii) computed the associated LS periodogram and 0.1\% FAP level, and iii) searched for significant peaks in the periodogram.  We considered the injected periodicity as successfully recovered if the strongest peak in the LS periodogram was above the 0.1\% FAP and if the relative error between the period marked by this peak and the injected period was $\le$5\%. Finally, we used these recovery rates to build sensitivity maps for amplitude and period as a function of magnitude and cadence.
Example sensitivity maps are shown in Figure \ref{maps}. Our simulations show that the minimum detectable variability amplitude at a fixed stellar magnitude is independent of period for periods in the 0.5--80 h range and cadences of either 2 min or 10 min (left and middle panels). The finding mostly holds for 30 min cadences too, except that the recovery rate deteriorates for periodicities shorter than about 1 h (right panels), especially for periods comparable to the 0.5 h sampling rate.

We use the detection rates in our simulated sensitivity maps of Figure~\ref{maps} to further assess  the minimum variability amplitude that we can detect for 0.5--80 h variability periods: considered as the lowest amplitude at which the recovery rate is still 95\%. We plot these amplitudes as a function of stellar magnitude in the top panel of Figure \ref{smr}. Our simulations show that at TESS magnitudes fainter than $\approx$16.5 we lose sensitivity to the typical $\le$1\% visible-wavelength variability of late-M to mid-L dwarfs \citep{2017MNRAS.472.2297M,2017ApJ...840...83M}. Figure \ref{smr} also compares these limits to the actual measured amplitudes for the variable targets in our survey (Table~\ref{mytable:target_details}). Four of our targets have amplitudes near the sensitivity limits for their magnitude. 

The bottom panel of Figure \ref{smr} shows the ratio between the minimum detectable amplitude (for 0.3-80 h periods) for a cadence of 10 min vs.\ 2 min and similarly for 30 min vs.\ 2 min. The amplitude ratio between 10 min and 2 min cadences is $\le$1.05 for the range 13-17.5 mag, but the ratio for 30 min and 2 min is significantly larger. The poorer sensitivity for data with cadences of 30 min is caused by less frequent sampling that blurs the photometric variability when stacking data every 30-min for short rotation periods. Given $>$1 h rotation periods for ultracool dwarfs (Section~\ref{sec:disc_periods}; see also \citealt{2021AJ....161..224T}) future searches for periodicities in ultra-cool dwarfs can be carried out by using the 10-min FFI from the extended mission of TESS without any additional need to obtain dedicated 2-min light curves.

\subsection{The period distribution of rapidly rotating ultra-cool dwarfs}
\label{sec:disc_periods}

Our survey targeted ultra-cool dwarfs with the fastest a priori known projected rotational velocities, but without prior determinations of photometric periods. By comparing with other photometric periods in the published literature, we are in position to assess the range of rotation periods in ultra-cool dwarfs. It is likely that our analysis is incomplete to $>$24 h periods, since outside of the TESS and Kepler missions that are mostly sensitive to $<$L0 dwarfs, long-duration monitoring of $\geq$L0 dwarfs has generally been limited to $<$24 h.

Our literature sample includes all previously confirmed photometric periods for dwarfs with spectral types later than M7, as known of this writing.  \citet{2021AJ....161..224T} recently analyzed a sample of 78 L, T, and Y dwarfs with known rotation periods and found that $\sim$1 h is a likely lower limit on the period for Jupiter-sized objects. Our expanded sample encompasses 128 ultra-cool dwarfs, including new L- and T-dwarf rotation periods presented in \citet{2022ApJ...924...68V} and 38 rotation period measurements for M7--M9.5 dwarfs presented in \citet{2017ApJ...834...85N,2017MNRAS.472.2297M,2021A&A...651L...7M,2022MNRAS.tmp..993A}, and this work. We verified the published late-M dwarf rotation periods using their TESS light curves, and analyzed them in a similar way as already presented in Sections \ref{sec:single} and \ref{sec:interference}. While the expanded sample of rotation periods no longer focuses only on the high-$v\sin i$ fast rotators targeted in our own observations, the overall distribution of periods is still biased to short periods. That is because most of the published periods of ultra-cool dwarfs are based on either $<$8 h ground-based or $\leq$24 h space-based monitoring campaigns.

In total we found TESS FFI data for 15 late-M dwarfs. We list their known rotation periods and our updated determinations from TESS data in Table \ref{mytable:results}. We plot all confirmed periods as a function of spectral type in Figure \ref{fig:rot}. In general, the TESS rotation periods are in good agreement with those reported in the literature. In most cases the TESS data refine the known rotation period, with the exception of TVLM 513-46, which is known to have a highly stable light curve from optical data covering $\sim$7 years. An interesting case is the binary LP 415-20 AB (M7+M9.5), which has a $4.4\pm1.6$ h photometric period from a ground-based measurement \citep{2017MNRAS.472.2297M}. However, an LS periodogram of the TESS data reveals two peaks at 3.6 h and 4.9 h (Fig.~\ref{lp415}). Using our beat pattern analysis from Section~\ref{sec:interference}, we arrive at accurate rotation periods for both components, which are in good agreement with the expected periods from combining radii and observed $v\sin i$. Four objects (LSR J0539+4038, LP 423-14, LSPM J1200+2048, LHS 2924) do not show any significant periodicity in the 27-days-long TESS data, which might indicate that either the targets are too faint for TESS to measure their photometric variability (we provide upper limits in Table \ref{mytable:results} using the limits in Fig. \ref{smr}) or the ground-based periodicities are spurious. 

The rotation periods of M7--Y0 dwarfs in Figure~\ref{fig:rot} show a large scatter between $\approx$1--24 h, with the greatest concentration in the 2--4 h range. 
The median rotation period is 3.9 hr, although this is likely skewed by the lack of sensitivity to potential $>$24-hour periods in photometric surveys, and by our focus on fast rotators within our own sample of 13 ultra-cool dwarf targets. All six of our periodically variable targets have periods between 2.4 h and 4.0 h. However, even if they were to be excluded, a significant enhancement of 2--4 h periods persists in Figure~\ref{fig:rot} (right panel).

A lower envelope to the period distribution is visible in the left panel of Figure~\ref{fig:rot} that runs from $\approx$2 h at spectral type M7.5--M8.5 (2MASS J01483864$-$3024396, TVLM~513$-$46; \citealt{2013MNRAS.428.2824K,2014ApJ...788...23W}) down to $\approx$1 h at spectral type T7 (2MASS J03480772-6022270; \citealt{2021AJ....161..224T}). The part of the envelope between M7--M9.5 effectively continues in an upsloping fashion toward warmer spectral types the trend observed in L--T dwarfs by \citet{2021AJ....161..224T}. We believe that the effect is real, and better sampled than the overall period distribution of M7--M9.5 dwarfs in Figure~\ref{fig:rot}, because of the specific focus of our photometric variability survey on ultra-cool dwarfs with high projected rotational velocities.

It is possible that the trend of an increasing minimum period toward warmer spectral subtypes in the M7--M9.5 is the result of structural stability considerations, as speculated in \citet{2021AJ....161..224T} for lower-mass brown dwarfs. The trend runs counter the expected \textit{decrease} in the minimum rotation period with increasing stellar mass at a constant final radius (as appropriate for these degenerate objects) from conservation of angular momentum. Thermonuclear energy generation, and correspondingly steeper temperature gradients and more vigorous convection in the interiors of very low-mass stars compared to substellar brown dwarfs, may drive the trend for longer minimum periods at higher masses.

\section{Conclusions}
\label{conc}
We combined ground-based $I$-band on 1~m-class telescopes and TESS data to search for photometric variability in a sample of 13 M7-L1.5 dwarfs, with $v\sin{i}>$30\,km\,s$^{-1}$. The ground-based data revealed periodicities in the 2 h to 3 h range for three targets (LP213-68AB, LP213-67, and J13365044+4751321), that we attribute to rotation as they are compatible with our estimations of rotation period based on the $v\sin{i}$ and the expected radii of our targets. Seven of our ground-based targets were also observed by TESS in either 2-min or 30-min cadence, and six of them (LP213-68AB, LP213-67, J13365044+4751321,  J14112131-2119503, J23515044-2537367, and LP859-1) show photometric variability compatible with rotation. While our ground-based data provide a better spatial resolution of the targets than TESS and are sensitive to $>$0.5\% $I$-band variability amplitudes, the long stare time provided by TESS allows exploration of a few hundred consecutive rotation periods, increasing the sensitivity to much smaller periodic photometric variability. 

We quantify the sensitivity of TESS to ultra-cool dwarf variability by simulating light curves with different amplitudes, periods and noise for TESS magnitudes in the range 13-19 mag. We find that TESS data (in either 2-min or 10-min cadence) can detect $\le$1\% photometric variability in ultra-cool dwarfs (such as typically seen in the red optical) and $\leq$80 h rotation periods for TESS magnitudes brighter than 16.5 with 95\% reliability. 

Finally, we compiled photometric periods for all $\ge$M7 dwarfs known to be photometrically variable from previous ground-based observations, and refined the rotation periods for 11 that have been observed by TESS. The entirety of the ultra-cool dwarf period distribution reveals a lower envelope on the rotation periods from $\approx$2 h for M7.5--M8.5 dwarfs to $\approx$1 h for late-T dwarfs. A larger, unbiased survey of photometric periods would be needed to confirm the trend, and its implications for the structural stability of rapidly rotating ultra-cool dwarfs.

\begin{table*}
    \small
    \centering
    \caption{Summary of measured periods.
        }
    \begin{tabular}{lccccccc}
\hline
\multicolumn{8}{c}{Ground-based}\\
\multicolumn{2}{l}{Target} & \multicolumn{1}{c}{SpT}& \multicolumn{1}{c}{TESS mag} &\multicolumn{2}{c}{A (\%)} & \multicolumn{2}{r}{$P_{\rm rot}$ (h)}\\
\hline
\multicolumn{2}{l}{LP 213-67} &\multicolumn{1}{c}{M7}&\multicolumn{1}{c}{13.549} &\multicolumn{2}{c}{0.64$\pm$0.05} & \multicolumn{2}{r}{$2.7\pm0.3$} \\
\multicolumn{2}{l}{LP 213-68AB} &\multicolumn{1}{c}{M8+L0}&\multicolumn{1}{c}{15.069} &\multicolumn{2}{c}{0.4$\pm$0.1} & \multicolumn{2}{r}{$2.3\pm0.3$} \\
\multicolumn{2}{l}{2MASS J13365044+4751321}&\multicolumn{1}{c}{M7} &\multicolumn{1}{c}{14.987} &\multicolumn{2}{c}{0.5$\pm$0.1} & \multicolumn{2}{r}{$2.6\pm0.2$} \\
\hline
\multicolumn{8}{c}{TESS}\\
\multicolumn{2}{l}{Target} & \multicolumn{1}{c}{SpT}& \multicolumn{1}{c}{TESS mag}& \multicolumn{1}{c}{Sector} & \multicolumn{1}{c}{A (\%)}& \multicolumn{1}{c}{$P_{\rm rot}$ (h)} & \multicolumn{1}{r}{$P_{\rm rot, adopted}^{\rm a}$ (h)}\\
\hline
\multicolumn{2}{l}{LP 213-67}& \multicolumn{1}{c}{M7} & \multicolumn{1}{c}{13.549}&\multicolumn{1}{c}{21} &\multicolumn{1}{c}{0.44$\pm$0.01} & \multicolumn{1}{c}{$3.0889\pm0.0002$} & \multicolumn{1}{r}{}\\
\multicolumn{2}{l}{LP 213-68AB}& \multicolumn{1}{c}{M8+L0} & \multicolumn{1}{c}{15.069}&\multicolumn{1}{c}{21} &\multicolumn{1}{c}{0.44$\pm$0.01} & \multicolumn{1}{c}{$2.5267\pm0.0002$} & \multicolumn{1}{r}{}\\
\multicolumn{2}{l}{2MASS J13365044+4751321}& \multicolumn{1}{c}{M7} & \multicolumn{1}{c}{14.987}&\multicolumn{1}{c}{16} &\multicolumn{1}{c}{0.55$\pm$0.03} & \multicolumn{1}{c}{$2.4219\pm0.0005$} & \multicolumn{1}{r}{$2.4221\pm0.0003$}\\
\multicolumn{2}{l}{} & \multicolumn{1}{c}{}& \multicolumn{1}{c}{}&\multicolumn{1}{c}{22} &\multicolumn{1}{c}{0.64$\pm$0.04} & \multicolumn{1}{c}{$2.4221\pm0.0003$} & \multicolumn{1}{r}{}\\
\multicolumn{2}{l}{} & \multicolumn{1}{c}{}& \multicolumn{1}{c}{}&\multicolumn{1}{c}{23} &\multicolumn{1}{c}{0.70$\pm$0.05} & \multicolumn{1}{c}{$2.4221\pm0.0004$} & \multicolumn{1}{r}{}\\
\multicolumn{2}{l}{2MASS J14112131-2119503}& \multicolumn{1}{c}{M7.5} & \multicolumn{1}{c}{14.905}&\multicolumn{1}{c}{11} &\multicolumn{1}{c}{0.42$\pm$0.07} & \multicolumn{1}{c}{$3.992\pm0.017$} & \multicolumn{1}{r}{}\\
\multicolumn{2}{l}{LP859-1} & \multicolumn{1}{c}{M7}& \multicolumn{1}{c}{14.454}&\multicolumn{1}{c}{11} &\multicolumn{1}{c}{0.21$\pm$0.07} & \multicolumn{1}{c}{$3.9597\pm0.0021$} & \multicolumn{1}{r}{$3.9614\pm0.0009$}\\
\multicolumn{2}{l}{} & \multicolumn{1}{c}{}& \multicolumn{1}{c}{}&\multicolumn{1}{c}{38} &\multicolumn{1}{c}{0.77$\pm$0.06} & \multicolumn{1}{c}{$3.9618\pm0.0010$} & \multicolumn{1}{r}{}\\
\multicolumn{2}{l}{2MASS J23515044-2537367}& \multicolumn{1}{c}{M9} & \multicolumn{1}{c}{15.608}&\multicolumn{1}{c}{2} &\multicolumn{1}{c}{0.44$\pm$0.03} & \multicolumn{1}{c}{$3.8630\pm0.0020$} & \multicolumn{1}{r}{$3.8632\pm0.0016$}\\
\multicolumn{2}{l}{} & \multicolumn{1}{c}{}& \multicolumn{1}{c}{}&\multicolumn{1}{c}{29} &\multicolumn{1}{c}{0.42$\pm$0.03} & \multicolumn{1}{c}{$3.8636\pm0.0025$} & \multicolumn{1}{r}{}\\
\hline
\multicolumn{8}{c}{Targets with previously reported periods from the literature recovered in the TESS FFI}\\
Target &SpT& TESS mag&$P_{\rm rot}$ (h)& Ref.&  Sector& $P_{\rm rot}$ (h)&A (\%) \\
           & &      &literature             &        &             & this work$^{\rm a}$
           & this work$^{\rm a}$ \\
\hline
2MASS J01483864-3024396	&M7.5		& 14.799	& 2--3 		&1  	&3, 30	&2.15012$\pm$0.00024			& 0.53$\pm$0.06\\
LP 944-20 				&M9.5 		&13.897	&2.5--3.7 		&2 	&4, 31	&3.8356 $\pm$0.0010			&0.17$\pm$0.04\\
LP 415-20$\rm ^b$			&M7+M9.5	&14.968	&4.4$\pm$1.6	&3 	&43, 44	&3.63515$\pm$0.00091$^{\rm c}$	&0.59$\pm$0.08 \\
						& 			& 		& 			&	&   		&4.9199$\pm$0.0035$^{\rm d}$	& 0.58$\pm$0.08\\
LSR J0539+4038 			& M8			& 13.501	& 15.6 		&4 	& 19 		&---				&  $<0.07^{\rm e}$\\
LP 423-14 				& M7 		&14.330	& 19.2		&4 	& 44 		& ---				&$<0.14^{\rm e}$\\
LHS 2397a 				& M8 		& 14.883 	& 5.9$\pm$0.5	&5  	& 9	 	& 5.984$\pm$0.001 				& 0.6$\pm$0.02\\
LSPM J1200+2048 			& M8 		& 15.300 	& 16.6  		&4 	& 22 		& --- 				&  $<0.34^{\rm e}$\\
LP 220-13 				&  M8 		& 14.346 	& 16.9  		&4 	&16, 23 	& 11.4334 $\pm$ 0.0058 			& 0.59$\pm$0.03\\
LSPM J1403+3007 			& M9 		& 15.53 	& 11.98  		&4	&23 		& 15.34$\pm$0.03 				& 0.66$\pm$0.06\\
LHS 2924	 				& M9 		& 15.100	& 1.8$\pm$0.2  & 5	& 23 		&	--- 			& $<0.28^{\rm e}$\\
TVLM 513-46 				& M8.5 		& 14.964 	& 1.959574$\pm$0.000002  &6		& 24 		& 1.95976$\pm$0.00043 & 0.43$\pm$0.08\\
2MASS J15210103+5053230 	& M7.5 		& 14.564 	& 3.1$\pm$0.9	& 3	& 16, 23, 24	& 3.0942$\pm$0.0010 		& 0.48$\pm$0.08\\
LSPM J1707+6439 			& M9 		& 15.492 	& 3.7$\pm$0.1  	& 7 		& 14--26, 40, 41 	& 3.62703$\pm$0.00031 	&0.49$\pm$0.1\\
LP 44-162 				& M7.5 		& 13.883 & 2.5--4 	& 3 	& 14--25, 40, 41	&3.89442 $\pm$ 0.00041 		& 0.18$\pm$0.04\\
LSR J1835+3259 			&M8.5 		& 13.286 	& 2.84$\pm$0.01  	& 8 		& 26, 40 			& 2.84140 $\pm$ 0.00039 & 0.34$\pm$0.03\\
\hline
\end{tabular}\\
 \begin{minipage}{175mm}
Notes: $\rm ^a$For targets observed in multiple sectors we analyze each sector separately. The adopted period (amplitude) is the weighted average of the rotation periods (amplitudes).\\
 $\rm ^b$ Light curve modeled as in Section \ref{sec:interference}.\\
 $\rm ^c$ Tentatively assigned to component A based on the observed $v$\,sin\,$i$.\\
 $\rm ^d$ Tentatively assigned to component B based on the observed $v$\,sin\,$i$.\\
 $\rm ^e$ TESS data do not show any significant periodicity. Upper limits are given based on the minimum amplitude detectable at a 95\% recovery rate for a cadence of 30-min from Figure \ref{smr}.\\
 References: 1-- \citet{2013MNRAS.428.2824K}; 2-- \citet{2006ApJ...644L..75M}; 3-- \citet{2017MNRAS.472.2297M}; 4-- \citet{2017ApJ...834...85N} ; 5-- \citet{1996ASPC..109..615M} ; 6-- \citet{2014ApJ...788...23W} ; 7-- \citet{2006A&A...448.1111R} ; 8-- \citet{2008ApJ...684..644H}.
\end{minipage}
\label{mytable:results}
\end{table*}
\section*{Acknowledgements}
TESS data was obtained from the Mikulski Archive for Space Telescopes (MAST). STScI is operated by the Association of Universities for Research in Astronomy, Inc., under NASA contract NAS5-26555. Support for MAST for non-HST data is provided by the NASA Office of Space Science via grant NNX13AC07G and by other grants and contracts. IRAF is distributed by the National Optical Astronomy Observatory, which is operated by the Association of Universities for Research in Astronomy (AURA) under a cooperative agreement with the National Science Foundation. 

\section*{Data Availability}
 
The TESS data underlying this article were obtained from the MAST data archive at the Space Telescope Science Institute (STScI). The ground-based data can be shared on reasonable request to the corresponding author.

\newpage



\bibliographystyle{mnras}
\bibliography{bibli} 

\begin{thebibliography}{}
\makeatletter
\relax
\def\mn@urlcharsother{\let\do\@makeother \do\$\do\&\do\#\do\^\do\_\do\%\do\~}
\def\mn@doi{\begingroup\mn@urlcharsother \@ifnextchar [ {\mn@doi@}
  {\mn@doi@[]}}
\def\mn@doi@[#1]#2{\def\@tempa{#1}\ifx\@tempa\@empty \href
  {http://dx.doi.org/#2} {doi:#2}\else \href {http://dx.doi.org/#2} {#1}\fi
  \endgroup}
\def\mn@eprint#1#2{\mn@eprint@#1:#2::\@nil}
\def\mn@eprint@arXiv#1{\href {http://arxiv.org/abs/#1} {{\tt arXiv:#1}}}
\def\mn@eprint@dblp#1{\href {http://dblp.uni-trier.de/rec/bibtex/#1.xml}
  {dblp:#1}}
\def\mn@eprint@#1:#2:#3:#4\@nil{\def\@tempa {#1}\def\@tempb {#2}\def\@tempc
  {#3}\ifx \@tempc \@empty \let \@tempc \@tempb \let \@tempb \@tempa \fi \ifx
  \@tempb \@empty \def\@tempb {arXiv}\fi \@ifundefined
  {mn@eprint@\@tempb}{\@tempb:\@tempc}{\expandafter \expandafter \csname
  mn@eprint@\@tempb\endcsname \expandafter{\@tempc}}}

\bibitem[\protect\citeauthoryear{{Andersson} et~al.,}{{Andersson}
  et~al.}{2022}]{2022MNRAS.tmp..993A}
{Andersson} A.,  et~al., 2022, \mn@doi [\mnras] {10.1093/mnras/stac1002}, \href
  {https://ui.adsabs.harvard.edu/abs/2022MNRAS.tmp..993A} {}

\bibitem[\protect\citeauthoryear{{Barclay}}{{Barclay}}{2017}]{2017zndo....888217B}
{Barclay} T.,  2017, {Tessgi/Ticgen: V1.0.0}, Zenodo,
  \mn@doi{10.5281/zenodo.888217}

\bibitem[\protect\citeauthoryear{{Bartlett} et~al.,}{{Bartlett}
  et~al.}{2017}]{2017AJ....154..151B}
{Bartlett} J.~L.,  et~al., 2017, \mn@doi [\aj] {10.3847/1538-3881/aa8457},
  \href {https://ui.adsabs.harvard.edu/abs/2017AJ....154..151B} {154, 151}

\bibitem[\protect\citeauthoryear{{Blake}, {Charbonneau}  \& {White}}{{Blake}
  et~al.}{2010}]{2010ApJ...723..684B}
{Blake} C.~H.,  {Charbonneau} D.,   {White} R.~J.,  2010, \mn@doi [\apj]
  {10.1088/0004-637X/723/1/684}, \href
  {https://ui.adsabs.harvard.edu/abs/2010ApJ...723..684B} {723, 684}

\bibitem[\protect\citeauthoryear{Bradley et~al.,}{Bradley
  et~al.}{2019}]{larry_bradley_2019_2533376}
Bradley L.,  et~al., 2019, astropy/photutils: v0.6,
  \mn@doi{10.5281/zenodo.2533376}, \url
  {https://doi.org/10.5281/zenodo.2533376}

\bibitem[\protect\citeauthoryear{{Buenzli}, {Apai}, {Radigan}, {Reid}  \&
  {Flateau}}{{Buenzli} et~al.}{2014}]{2014ApJ...782...77B}
{Buenzli} E.,  {Apai} D.,  {Radigan} J.,  {Reid} I.~N.,   {Flateau} D.,  2014,
  \mn@doi [\apj] {10.1088/0004-637X/782/2/77}, \href
  {http://adsabs.harvard.edu/abs/2014ApJ...782...77B} {782, 77}

\bibitem[\protect\citeauthoryear{{Chabrier} \& {Baraffe}}{{Chabrier} \&
  {Baraffe}}{2000}]{2000ARA&A..38..337C}
{Chabrier} G.,  {Baraffe} I.,  2000, \mn@doi [\araa]
  {10.1146/annurev.astro.38.1.337}, \href
  {https://ui.adsabs.harvard.edu/abs/2000ARA%26A..38..337C} {38, 337}

\bibitem[\protect\citeauthoryear{{Close}, {Siegler}, {Freed}  \&
  {Biller}}{{Close} et~al.}{2003}]{2003ApJ...587..407C}
{Close} L.~M.,  {Siegler} N.,  {Freed} M.,   {Biller} B.,  2003, \mn@doi [\apj]
  {10.1086/368177}, \href
  {https://ui.adsabs.harvard.edu/abs/2003ApJ...587..407C} {587, 407}

\bibitem[\protect\citeauthoryear{{Cushing} et~al.,}{{Cushing}
  et~al.}{2016}]{2016ApJ...823..152C}
{Cushing} M.~C.,  et~al., 2016, \mn@doi [\apj] {10.3847/0004-637X/823/2/152},
  \href {https://ui.adsabs.harvard.edu/abs/2016ApJ...823..152C} {823, 152}

\bibitem[\protect\citeauthoryear{{Dupuy} \& {Liu}}{{Dupuy} \&
  {Liu}}{2017}]{2017ApJS..231...15D}
{Dupuy} T.~J.,  {Liu} M.~C.,  2017, \mn@doi [\apjs] {10.3847/1538-4365/aa5e4c},
  \href {https://ui.adsabs.harvard.edu/abs/2017ApJS..231...15D} {231, 15}

\bibitem[\protect\citeauthoryear{{Feinstein} et~al.,}{{Feinstein}
  et~al.}{2019}]{2019PASP..131i4502F}
{Feinstein} A.~D.,  et~al., 2019, \mn@doi [\pasp] {10.1088/1538-3873/ab291c},
  \href {https://ui.adsabs.harvard.edu/abs/2019PASP..131i4502F} {131, 094502}

\bibitem[\protect\citeauthoryear{{Filippazzo}, {Rice}, {Faherty}, {Cruz}, {Van
  Gordon}  \& {Looper}}{{Filippazzo} et~al.}{2015}]{2015ApJ...810..158F}
{Filippazzo} J.~C.,  {Rice} E.~L.,  {Faherty} J.,  {Cruz} K.~L.,  {Van Gordon}
  M.~M.,   {Looper} D.~L.,  2015, \mn@doi [\apj] {10.1088/0004-637X/810/2/158},
  \href {https://ui.adsabs.harvard.edu/abs/2015ApJ...810..158F} {810, 158}

\bibitem[\protect\citeauthoryear{{Foreman-Mackey}, {Hogg}, {Lang}  \&
  {Goodman}}{{Foreman-Mackey} et~al.}{2013}]{2013PASP..125..306F}
{Foreman-Mackey} D.,  {Hogg} D.~W.,  {Lang} D.,   {Goodman} J.,  2013, \mn@doi
  [\pasp] {10.1086/670067}, \href
  {https://ui.adsabs.harvard.edu/abs/2013PASP..125..306F} {125, 306}

\bibitem[\protect\citeauthoryear{{Foreman-Mackey}, {Agol}, {Ambikasaran}  \&
  {Angus}}{{Foreman-Mackey} et~al.}{2017}]{2017AJ....154..220F}
{Foreman-Mackey} D.,  {Agol} E.,  {Ambikasaran} S.,   {Angus} R.,  2017,
  \mn@doi [\aj] {10.3847/1538-3881/aa9332}, \href
  {https://ui.adsabs.harvard.edu/abs/2017AJ....154..220F} {154, 220}

\bibitem[\protect\citeauthoryear{{Gagn{\'e}} et~al.,}{{Gagn{\'e}}
  et~al.}{2018}]{2018ApJ...856...23G}
{Gagn{\'e}} J.,  et~al., 2018, \mn@doi [\apj] {10.3847/1538-4357/aaae09}, \href
  {https://ui.adsabs.harvard.edu/abs/2018ApJ...856...23G} {856, 23}

\bibitem[\protect\citeauthoryear{{Gizis}, {Monet}, {Reid}, {Kirkpatrick}  \&
  {Burgasser}}{{Gizis} et~al.}{2000}]{2000MNRAS.311..385G}
{Gizis} J.~E.,  {Monet} D.~G.,  {Reid} I.~N.,  {Kirkpatrick} J.~D.,
  {Burgasser} A.~J.,  2000, \mn@doi [\mnras]
  {10.1046/j.1365-8711.2000.03060.x}, \href
  {https://ui.adsabs.harvard.edu/abs/2000MNRAS.311..385G} {311, 385}

\bibitem[\protect\citeauthoryear{{Gizis}, {Burgasser}, {Berger}, {Williams},
  {Vrba}, {Cruz}  \& {Metchev}}{{Gizis} et~al.}{2013}]{2013ApJ...779..172G}
{Gizis} J.~E.,  {Burgasser} A.~J.,  {Berger} E.,  {Williams} P.~K.~G.,  {Vrba}
  F.~J.,  {Cruz} K.~L.,   {Metchev} S.,  2013, \mn@doi [\apj]
  {10.1088/0004-637X/779/2/172}, \href
  {https://ui.adsabs.harvard.edu/abs/2013ApJ...779..172G} {779, 172}

\bibitem[\protect\citeauthoryear{{Goodman} \& {Weare}}{{Goodman} \&
  {Weare}}{2010}]{2010CAMCS...5...65G}
{Goodman} J.,  {Weare} J.,  2010, \mn@doi [Communications in Applied
  Mathematics and Computational Science] {10.2140/camcos.2010.5.65}, \href
  {https://ui.adsabs.harvard.edu/abs/2010CAMCS...5...65G} {5, 65}

\bibitem[\protect\citeauthoryear{{Hallinan}, {Antonova}, {Doyle}, {Bourke},
  {Lane}  \& {Golden}}{{Hallinan} et~al.}{2008}]{2008ApJ...684..644H}
{Hallinan} G.,  {Antonova} A.,  {Doyle} J.~G.,  {Bourke} S.,  {Lane} C.,
  {Golden} A.,  2008, \mn@doi [\apj] {10.1086/590360}, \href
  {https://ui.adsabs.harvard.edu/abs/2008ApJ...684..644H} {684, 644}

\bibitem[\protect\citeauthoryear{{Harding}, {Hallinan}, {Boyle}, {Golden},
  {Singh}, {Sheehan}, {Zavala}  \& {Butler}}{{Harding}
  et~al.}{2013}]{2013ApJ...779..101H}
{Harding} L.~K.,  {Hallinan} G.,  {Boyle} R.~P.,  {Golden} A.,  {Singh} N.,
  {Sheehan} B.,  {Zavala} R.~T.,   {Butler} R.~F.,  2013, \mn@doi [\apj]
  {10.1088/0004-637X/779/2/101}, \href
  {http://adsabs.harvard.edu/abs/2013ApJ...779..101H} {779, 101}

\bibitem[\protect\citeauthoryear{{Hooten} \& {Hall}}{{Hooten} \&
  {Hall}}{1990}]{1990ApJS...74..225H}
{Hooten} J.~T.,  {Hall} D.~S.,  1990, \mn@doi [\apjs] {10.1086/191497}, \href
  {https://ui.adsabs.harvard.edu/abs/1990ApJS...74..225H} {74, 225}

\bibitem[\protect\citeauthoryear{{Ivezi{\'c}}, {Connelly}, {VanderPlas}  \&
  {Gray}}{{Ivezi{\'c}} et~al.}{2014}]{2014sdmm.book.....I}
{Ivezi{\'c}} {\v{Z}}.,  {Connelly} A.~J.,  {VanderPlas} J.~T.,   {Gray} A.,
  2014, {Statistics, Data Mining, and Machine Learning in Astronomy}

\bibitem[\protect\citeauthoryear{{James}}{{James}}{1964}]{1964ApJ...140..552J}
{James} R.~A.,  1964, \mn@doi [\apj] {10.1086/147949}, \href
  {https://ui.adsabs.harvard.edu/abs/1964ApJ...140..552J} {140, 552}

\bibitem[\protect\citeauthoryear{{Jenkins}, {Ramsey}, {Jones}, {Pavlenko},
  {Gallardo}, {Barnes}  \& {Pinfield}}{{Jenkins}
  et~al.}{2009}]{2009ApJ...704..975J}
{Jenkins} J.~S.,  {Ramsey} L.~W.,  {Jones} H.~R.~A.,  {Pavlenko} Y.,
  {Gallardo} J.,  {Barnes} J.~R.,   {Pinfield} D.~J.,  2009, \mn@doi [\apj]
  {10.1088/0004-637X/704/2/975}, \href
  {https://ui.adsabs.harvard.edu/abs/2009ApJ...704..975J} {704, 975}

\bibitem[\protect\citeauthoryear{{Koen}}{{Koen}}{2005}]{2005MNRAS.360.1132K}
{Koen} C.,  2005, \mn@doi [\mnras] {10.1111/j.1365-2966.2005.09119.x}, \href
  {https://ui.adsabs.harvard.edu/abs/2005MNRAS.360.1132K} {360, 1132}

\bibitem[\protect\citeauthoryear{{Koen}}{{Koen}}{2013}]{2013MNRAS.428.2824K}
{Koen} C.,  2013, \mn@doi [\mnras] {10.1093/mnras/sts208}, \href
  {https://ui.adsabs.harvard.edu/abs/2013MNRAS.428.2824K} {428, 2824}

\bibitem[\protect\citeauthoryear{{Littlefair}, {Burningham}  \&
  {Helling}}{{Littlefair} et~al.}{2017}]{2017MNRAS.466.4250L}
{Littlefair} S.~P.,  {Burningham} B.,   {Helling} C.,  2017, \mn@doi [\mnras]
  {10.1093/mnras/stw3376}, \href
  {https://ui.adsabs.harvard.edu/abs/2017MNRAS.466.4250L} {466, 4250}

\bibitem[\protect\citeauthoryear{{Liu}, {Dupuy}  \& {Allers}}{{Liu}
  et~al.}{2016}]{2016ApJ...833...96L}
{Liu} M.~C.,  {Dupuy} T.~J.,   {Allers} K.~N.,  2016, \mn@doi [\apj]
  {10.3847/1538-4357/833/1/96}, \href
  {https://ui.adsabs.harvard.edu/abs/2016ApJ...833...96L} {833, 96}

\bibitem[\protect\citeauthoryear{{Lomb}}{{Lomb}}{1976}]{1976Ap&SS..39..447L}
{Lomb} N.~R.,  1976, \mn@doi [\apss] {10.1007/BF00648343}, \href
  {https://ui.adsabs.harvard.edu/abs/1976Ap%26SS..39..447L} {39, 447}

\bibitem[\protect\citeauthoryear{{Martin} \& {Zapatero-Osorio}}{{Martin} \&
  {Zapatero-Osorio}}{1997}]{1997MNRAS.286L..17M}
{Martin} E.~L.,  {Zapatero-Osorio} M.~R.,  1997, \mn@doi [\mnras]
  {10.1093/mnras/286.1.L17}, \href
  {https://ui.adsabs.harvard.edu/abs/1997MNRAS.286L..17M} {286, L17}

\bibitem[\protect\citeauthoryear{{Martin}, {Zapatero Osorio}  \&
  {Rebolo}}{{Martin} et~al.}{1996}]{1996ASPC..109..615M}
{Martin} E.~L.,  {Zapatero Osorio} M.~R.,   {Rebolo} R.,  1996, in
  {Pallavicini} R.,  {Dupree} A.~K.,  eds,  Astronomical Society of the Pacific
  Conference Series Vol. 109, Cool Stars, Stellar Systems, and the Sun. p.~615

\bibitem[\protect\citeauthoryear{{Mart{\'\i}n}, {Guenther}, {Zapatero Osorio},
  {Bouy}  \& {Wainscoat}}{{Mart{\'\i}n} et~al.}{2006}]{2006ApJ...644L..75M}
{Mart{\'\i}n} E.~L.,  {Guenther} E.,  {Zapatero Osorio} M.~R.,  {Bouy} H.,
  {Wainscoat} R.,  2006, \mn@doi [\apjl] {10.1086/505343}, \href
  {https://ui.adsabs.harvard.edu/abs/2006ApJ...644L..75M} {644, L75}

\bibitem[\protect\citeauthoryear{{Martin} et~al.,}{{Martin}
  et~al.}{2017}]{2017ApJ...838...73M}
{Martin} E.~C.,  et~al., 2017, \mn@doi [\apj] {10.3847/1538-4357/aa6338}, \href
  {https://ui.adsabs.harvard.edu/abs/2017ApJ...838...73M} {838, 73}

\bibitem[\protect\citeauthoryear{{Metchev} et~al.,}{{Metchev}
  et~al.}{2015}]{2015ApJ...799..154M}
{Metchev} S.~A.,  et~al., 2015, \mn@doi [\apj] {10.1088/0004-637X/799/2/154},
  \href {http://adsabs.harvard.edu/abs/2015ApJ...799..154M} {799, 154}

\bibitem[\protect\citeauthoryear{{Miles-P{\'a}ez}}{{Miles-P{\'a}ez}}{2021}]{2021A&A...651L...7M}
{Miles-P{\'a}ez} P.~A.,  2021, \mn@doi [\aap] {10.1051/0004-6361/202141203},
  \href {https://ui.adsabs.harvard.edu/abs/2021A&A...651L...7M} {651, L7}

\bibitem[\protect\citeauthoryear{{Miles-P{\'a}ez}, {Pall{\'e}}  \& {Zapatero
  Osorio}}{{Miles-P{\'a}ez} et~al.}{2017a}]{2017MNRAS.472.2297M}
{Miles-P{\'a}ez} P.~A.,  {Pall{\'e}} E.,   {Zapatero Osorio} M.~R.,  2017a,
  \mn@doi [\mnras] {10.1093/mnras/stx2191}, \href
  {http://adsabs.harvard.edu/abs/2017MNRAS.472.2297M} {472, 2297}

\bibitem[\protect\citeauthoryear{{Miles-P{\'a}ez}, {Metchev}, {Heinze}  \&
  {Apai}}{{Miles-P{\'a}ez} et~al.}{2017b}]{2017ApJ...840...83M}
{Miles-P{\'a}ez} P.~A.,  {Metchev} S.~A.,  {Heinze} A.,   {Apai} D.,  2017b,
  \mn@doi [\apj] {10.3847/1538-4357/aa6f11}, \href
  {https://ui.adsabs.harvard.edu/abs/2017ApJ...840...83M} {840, 83}

\bibitem[\protect\citeauthoryear{{Miles-P{\'a}ez} et~al.,}{{Miles-P{\'a}ez}
  et~al.}{2019}]{2019ApJ...883..181M}
{Miles-P{\'a}ez} P.~A.,  et~al., 2019, \mn@doi [\apj]
  {10.3847/1538-4357/ab3d25}, \href
  {https://ui.adsabs.harvard.edu/abs/2019ApJ...883..181M} {883, 181}

\bibitem[\protect\citeauthoryear{{Mohanty} \& {Basri}}{{Mohanty} \&
  {Basri}}{2003}]{2003ApJ...583..451M}
{Mohanty} S.,  {Basri} G.,  2003, \mn@doi [\apj] {10.1086/345097}, \href
  {https://ui.adsabs.harvard.edu/abs/2003ApJ...583..451M} {583, 451}

\bibitem[\protect\citeauthoryear{{Newton}, {Irwin}, {Charbonneau}, {Berlind},
  {Calkins}  \& {Mink}}{{Newton} et~al.}{2017}]{2017ApJ...834...85N}
{Newton} E.~R.,  {Irwin} J.,  {Charbonneau} D.,  {Berlind} P.,  {Calkins}
  M.~L.,   {Mink} J.,  2017, \mn@doi [\apj] {10.3847/1538-4357/834/1/85}, \href
  {https://ui.adsabs.harvard.edu/abs/2017ApJ...834...85N} {834, 85}

\bibitem[\protect\citeauthoryear{{Radigan}, {Lafreni{\`e}re}, {Jayawardhana}
  \& {Artigau}}{{Radigan} et~al.}{2014}]{2014ApJ...793...75R}
{Radigan} J.,  {Lafreni{\`e}re} D.,  {Jayawardhana} R.,   {Artigau} E.,  2014,
  \mn@doi [\apj] {10.1088/0004-637X/793/2/75}, \href
  {https://ui.adsabs.harvard.edu/abs/2014ApJ...793...75R} {793, 75}

\bibitem[\protect\citeauthoryear{{Rasmussen} \& {Williams}}{{Rasmussen} \&
  {Williams}}{2006}]{2006gpml.book.....R}
{Rasmussen} C.~E.,  {Williams} C. K.~I.,  2006, {Gaussian Processes for Machine
  Learning}

\bibitem[\protect\citeauthoryear{{Reid}, {Lewitus}, {Allen}, {Cruz}  \&
  {Burgasser}}{{Reid} et~al.}{2006}]{2006AJ....132..891R}
{Reid} I.~N.,  {Lewitus} E.,  {Allen} P.~R.,  {Cruz} K.~L.,   {Burgasser}
  A.~J.,  2006, \mn@doi [\aj] {10.1086/505626}, \href
  {https://ui.adsabs.harvard.edu/abs/2006AJ....132..891R} {132, 891}

\bibitem[\protect\citeauthoryear{{Reiners} \& {Basri}}{{Reiners} \&
  {Basri}}{2008}]{2008ApJ...684.1390R}
{Reiners} A.,  {Basri} G.,  2008, \mn@doi [\apj] {10.1086/590073}, \href
  {https://ui.adsabs.harvard.edu/abs/2008ApJ...684.1390R} {684, 1390}

\bibitem[\protect\citeauthoryear{{Reiners}, {Bean}, {Huber}, {Dreizler},
  {Seifahrt}  \& {Czesla}}{{Reiners} et~al.}{2010}]{2010ApJ...710..432R}
{Reiners} A.,  {Bean} J.~L.,  {Huber} K.~F.,  {Dreizler} S.,  {Seifahrt} A.,
  {Czesla} S.,  2010, \mn@doi [\apj] {10.1088/0004-637X/710/1/432}, \href
  {https://ui.adsabs.harvard.edu/abs/2010ApJ...710..432R} {710, 432}

\bibitem[\protect\citeauthoryear{{Rice}, {Barman}, {Mclean}, {Prato}  \&
  {Kirkpatrick}}{{Rice} et~al.}{2010}]{2010ApJS..186...63R}
{Rice} E.~L.,  {Barman} T.,  {Mclean} I.~S.,  {Prato} L.,   {Kirkpatrick}
  J.~D.,  2010, \mn@doi [\apjs] {10.1088/0067-0049/186/1/63}, \href
  {https://ui.adsabs.harvard.edu/abs/2010ApJS..186...63R} {186, 63}

\bibitem[\protect\citeauthoryear{{Ricker} et~al.,}{{Ricker}
  et~al.}{2014}]{2014SPIE.9143E..20R}
{Ricker} G.~R.,  et~al., 2014, in {Oschmann} Jacobus~M. J.,  {Clampin} M.,
  {Fazio} G.~G.,   {MacEwen} H.~A.,  eds,  Society of Photo-Optical
  Instrumentation Engineers (SPIE) Conference Series Vol. 9143, Space
  Telescopes and Instrumentation 2014: Optical, Infrared, and Millimeter Wave.
  p. 914320 (\mn@eprint {arXiv} {1406.0151}), \mn@doi{10.1117/12.2063489}

\bibitem[\protect\citeauthoryear{{Roberts}, {Osborne}, {Ebden}, {Reece},
  {Gibson}  \& {Aigrain}}{{Roberts} et~al.}{2012}]{2012RSPTA.37110550R}
{Roberts} S.,  {Osborne} M.,  {Ebden} M.,  {Reece} S.,  {Gibson} N.,
  {Aigrain} S.,  2012, \mn@doi [Philosophical Transactions of the Royal Society
  of London Series A] {10.1098/rsta.2011.0550}, \href
  {https://ui.adsabs.harvard.edu/abs/2012RSPTA.37110550R} {371, 20110550}

\bibitem[\protect\citeauthoryear{{Rockenfeller}, {Bailer-Jones}  \&
  {Mundt}}{{Rockenfeller} et~al.}{2006}]{2006A&A...448.1111R}
{Rockenfeller} B.,  {Bailer-Jones} C.~A.~L.,   {Mundt} R.,  2006, \mn@doi
  [\aap] {10.1051/0004-6361:20054150}, \href
  {https://ui.adsabs.harvard.edu/abs/2006A%26A...448.1111R} {448, 1111}

\bibitem[\protect\citeauthoryear{{Scargle}}{{Scargle}}{1982}]{1982ApJ...263..835S}
{Scargle} J.~D.,  1982, \mn@doi [\apj] {10.1086/160554}, \href
  {http://adsabs.harvard.edu/abs/1982ApJ...263..835S} {263, 835}

\bibitem[\protect\citeauthoryear{{Schwarz}}{{Schwarz}}{1978}]{1978AnSta...6..461S}
{Schwarz} G.,  1978, Annals of Statistics, \href
  {https://ui.adsabs.harvard.edu/abs/1978AnSta...6..461S} {6, 461}

\bibitem[\protect\citeauthoryear{{Siegler}, {Close}, {Cruz}, {Mart{\'{\i}}n}
  \& {Reid}}{{Siegler} et~al.}{2005}]{2005ApJ...621.1023S}
{Siegler} N.,  {Close} L.~M.,  {Cruz} K.~L.,  {Mart{\'{\i}}n} E.~L.,   {Reid}
  I.~N.,  2005, \mn@doi [\apj] {10.1086/427743}, \href
  {https://ui.adsabs.harvard.edu/abs/2005ApJ...621.1023S} {621, 1023}

\bibitem[\protect\citeauthoryear{{Skumanich}}{{Skumanich}}{1972}]{1972ApJ...171..565S}
{Skumanich} A.,  1972, \mn@doi [\apj] {10.1086/151310}, \href
  {https://ui.adsabs.harvard.edu/abs/1972ApJ...171..565S} {171, 565}

\bibitem[\protect\citeauthoryear{{Stassun} et~al.,}{{Stassun}
  et~al.}{2018}]{2018AJ....156..102S}
{Stassun} K.~G.,  et~al., 2018, \mn@doi [\aj] {10.3847/1538-3881/aad050}, \href
  {https://ui.adsabs.harvard.edu/abs/2018AJ....156..102S} {156, 102}

\bibitem[\protect\citeauthoryear{{Tannock} et~al.,}{{Tannock}
  et~al.}{2021}]{2021AJ....161..224T}
{Tannock} M.~E.,  et~al., 2021, \mn@doi [\aj] {10.3847/1538-3881/abeb67}, \href
  {https://ui.adsabs.harvard.edu/abs/2021AJ....161..224T} {161, 224}

\bibitem[\protect\citeauthoryear{Taylor}{Taylor}{1990}]{taylor1990interpretation}
Taylor R.,  1990, JDMS, 6, 35

\bibitem[\protect\citeauthoryear{{Tinney} \& {Tolley}}{{Tinney} \&
  {Tolley}}{1999}]{1999MNRAS.304..119T}
{Tinney} C.~G.,  {Tolley} A.~J.,  1999, \mn@doi [\mnras]
  {10.1046/j.1365-8711.1999.02297.x}, \href
  {https://ui.adsabs.harvard.edu/abs/1999MNRAS.304..119T} {304, 119}

\bibitem[\protect\citeauthoryear{{Tremblin}, {Amundsen}, {Mourier}, {Baraffe},
  {Chabrier}, {Drummond}, {Homeier}  \& {Venot}}{{Tremblin}
  et~al.}{2015}]{2015ApJ...804L..17T}
{Tremblin} P.,  {Amundsen} D.~S.,  {Mourier} P.,  {Baraffe} I.,  {Chabrier} G.,
   {Drummond} B.,  {Homeier} D.,   {Venot} O.,  2015, \mn@doi [\apjl]
  {10.1088/2041-8205/804/1/L17}, \href
  {https://ui.adsabs.harvard.edu/abs/2015ApJ...804L..17T} {804, L17}

\bibitem[\protect\citeauthoryear{{Tsuji}, {Ohnaka}, {Aoki}  \&
  {Nakajima}}{{Tsuji} et~al.}{1996}]{1996A&A...308L..29T}
{Tsuji} T.,  {Ohnaka} K.,  {Aoki} W.,   {Nakajima} T.,  1996, \aap, \href
  {https://ui.adsabs.harvard.edu/abs/1996A&A...308L..29T} {308, L29}

\bibitem[\protect\citeauthoryear{{VanderPlas}}{{VanderPlas}}{2018}]{2018ApJS..236...16V}
{VanderPlas} J.~T.,  2018, \mn@doi [\apjs] {10.3847/1538-4365/aab766}, \href
  {http://adsabs.harvard.edu/abs/2018ApJS..236...16V} {236, 16}

\bibitem[\protect\citeauthoryear{{Vanderspek}}{{Vanderspek}}{2019}]{2019ESS.....433312V}
{Vanderspek} R.,  2019, in AAS/Division for Extreme Solar Systems Abstracts. p.
  333.12

\bibitem[\protect\citeauthoryear{{Vos}, {Faherty}, {Gagn{\'e}}, {Marley},
  {Metchev}, {Gizis}, {Rice}  \& {Cruz}}{{Vos}
  et~al.}{2022}]{2022ApJ...924...68V}
{Vos} J.~M.,  {Faherty} J.~K.,  {Gagn{\'e}} J.,  {Marley} M.,  {Metchev} S.,
  {Gizis} J.,  {Rice} E.~L.,   {Cruz} K.,  2022, \mn@doi [\apj]
  {10.3847/1538-4357/ac4502}, \href
  {https://ui.adsabs.harvard.edu/abs/2022ApJ...924...68V} {924, 68}

\bibitem[\protect\citeauthoryear{{Wolszczan} \& {Route}}{{Wolszczan} \&
  {Route}}{2014}]{2014ApJ...788...23W}
{Wolszczan} A.,  {Route} M.,  2014, \mn@doi [\apj]
  {10.1088/0004-637X/788/1/23}, \href
  {http://adsabs.harvard.edu/abs/2014ApJ...788...23W} {788, 23}

\makeatother
\end{thebibliography}

\appendix

\section{Additional figures}

\label{app1}
\begin{centering}
\begin{figure*}
\includegraphics[width=0.99\textwidth]{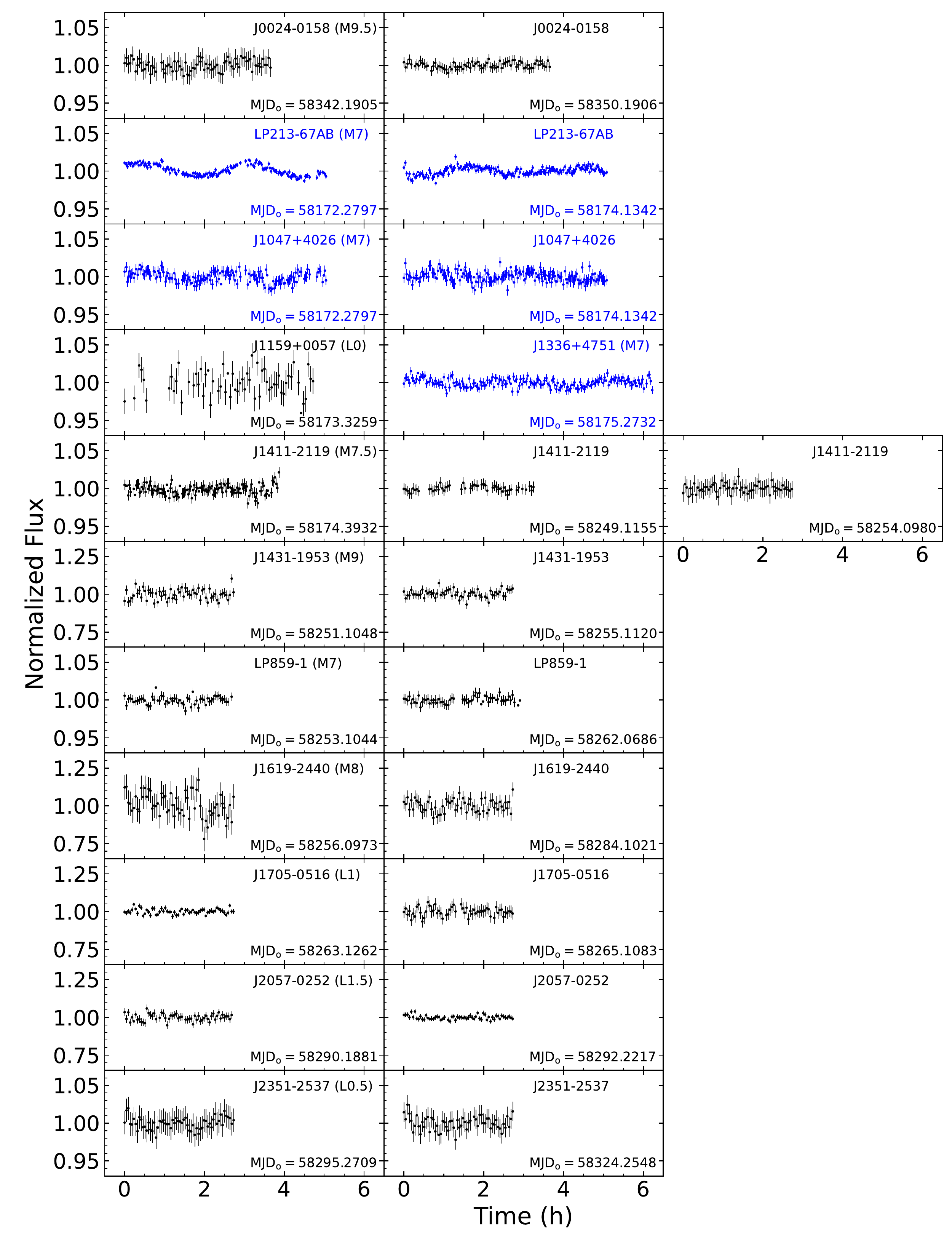}
\caption{Differential photometry for our sample in each epoch of observations. MJD's indicate the start time of observations. The different columns show observations on multiple epochs for most targets, except for J1159+0057 and J1336+4751, which were observed on only one night each and are shown in row 4. Real variability can be visually observed in LP213-67, LP213-68AB, and J1336+4751, which are plotted with blue color.}
\label{myfig:full_lc}
\end{figure*}
\end{centering}

\begin{centering}
\begin{figure*}
\includegraphics[width=0.99\textwidth]{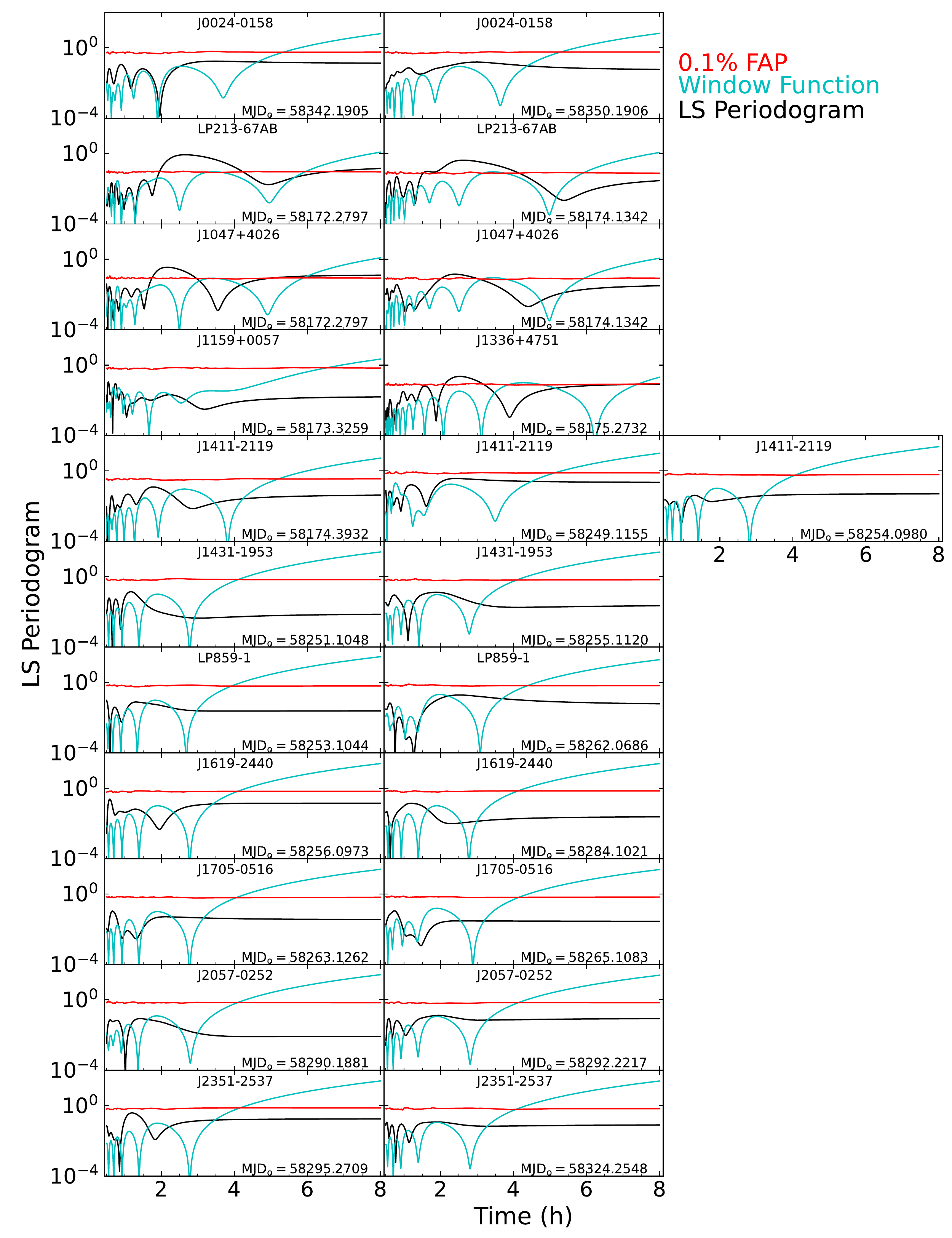}
\caption{LS periodograms for each lightcurve (black) shown in Figure \ref{myfig:full_lc}. A 0.1\% False-Alarm-Probability is shown in red, which was computed by bootstrapping using $10^4$ samples. Any peak above the FAP line has a confidence of 99.9\%. The window function of each periodogram is also shown in cyan. As in Figure~\ref{myfig:full_lc}, different columns show observations on multiple epochs for most targets, except for J1159+0057 and J1336+4751, which were observed on only one night each and are shown in row 4.}
\label{myfig:full_ls}
\end{figure*}
\end{centering}

\begin{centering}
\begin{figure}
\includegraphics[width=0.49\textwidth]{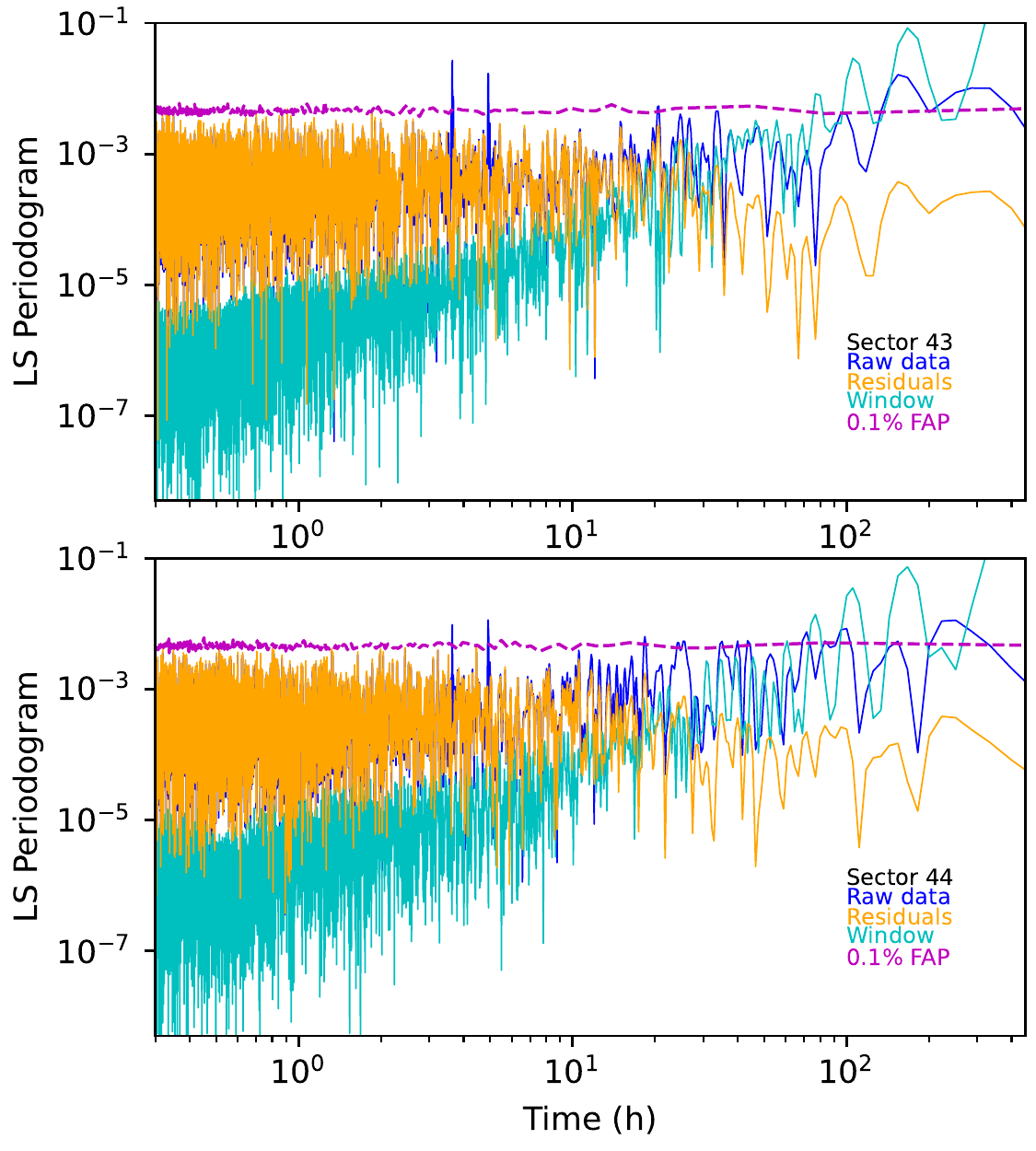}
\caption{LS periodogram for the raw data (blue) of LP415-20AB observed in Sectors 43 (top) and 44 (bottom). The orange line in both panels shows the LS periodogram of the residuals of the data in each sector (i.e., after fitting and subtracting from the data the two-periods model described in Sec. \ref{sec:interference}). The window function (cyan) and the 99.9\% confidence level (magenta) for each data set are also shown. Data in both sectors clearly show two significant peaks at 3.6 h and 4.9 h that we attribute to the rotation periods of the individual components of the binary.}
\label{lp415}
\end{figure}
\end{centering}



\bsp	
\label{lastpage}
\end{document}